\documentclass[aps,prb,twocolumn,superscriptaddress,showpacs,amsmath,amssymb]{revtex4-2}
\usepackage{graphicx}
\usepackage{dcolumn}
\usepackage{bm}
\usepackage{color}
\usepackage{subfig}
\usepackage{verbatim}
\usepackage{ragged2e}
\DeclareCaptionJustification{justified}{\justifying}
\renewcommand\l{\langle}
\renewcommand\r{\rangle}
\renewcommand\Re{\operatorname{Re}}
\renewcommand\Im{\operatorname{Im}}
\newcommand{\U}{\hat{\mathcal U}}
\renewcommand{\H}{\hat{\mathcal H}_m}

\newcommand{\SOC}{\lambda}
\newcommand{\Hs}{\hat{\mathcal H}_{J}}
\newcommand{\norma}{\mathcal N}
\newcommand{\J}{\mathcal J_{\text{ex}}}
\newcommand{\Jx}{\mathcal J_0}
\newcommand{\DD}{6\Delta}
\newcommand{\D}{3\Delta}
\newcommand{\omegaj}{\omega_{\mathcal J}}
\renewcommand{\P}{\Psi^{(1,2)}_{\text g}}
\newcommand{\Hcrx}{\hat{\mathcal H}_{\text{cr}x}}
\newcommand{\Hcrz}{\hat{\mathcal H}_{\text{cr}z}}
\newcommand{\Hcr}{\hat{\mathcal H}_{\text{cr}}}
\newcommand{\lf}{\left}
\newcommand{\rt}{\right}
\newcommand{\la}{\langle}
\newcommand{\ra}{\rangle}

\begin{document}
\title{Heisenberg representation of nonthermal ultrafast laser excitation of magnetic precessions}
\author{Daria Popova-Gorelova } 
\affiliation{
Peter Gr\"unberg Institute and Institute for Advanced Simulation,
Forschungszentrum J\"ulich and JARA, 52425 J\"ulich,
Germany }
\affiliation{Department of Physics, Universit\"at Hamburg, Notkestrasse 9, D-22607 Hamburg, Germany} 
\affiliation{The Hamburg Centre for Ultrafast Imaging (CUI), Luruper Chaussee 149, 22607 Hamburg, Germany}
\author{Andreas Bringer}
\affiliation{
Peter Gr\"unberg Institute and Institute for Advanced Simulation,
Forschungszentrum J\"ulich and JARA, 52425 J\"ulich,
Germany }
\author{Stefan Bl\"ugel}
\affiliation{
Peter Gr\"unberg Institute and Institute for Advanced Simulation,
Forschungszentrum J\"ulich and JARA, 52425 J\"ulich,
Germany }
\date{\today}
\begin{abstract}

We derive the Heisenberg representation of the ultrafast inverse Faraday effect that provides the time evolution of  magnetic vectors of a magnetic system during its interaction with a laser pulse. We obtain a time-dependent effective magnetic operator acting in the Hilbert space of the total angular momentum that describes a process of nonthermal excitation of magnetic precessions in an electronic system by a circularly polarized laser pulse. The magnetic operator separates the effect of the laser pulse on the magnetic system from other magnetic interactions. The effective magnetic operator provides the equations of motion of magnetic vectors during the excitation by the laser. We show that magnetization dynamics calculated with these equations is equivalent to magnetization dynamics calculated with the time-dependent Schr\"odinger equation, which takes into account the interaction of an electronic system with the electric field of light. We model and compare laser-induced precessions of magnetic sublattices of an easy-plane and an easy-axis antiferromagnetic systems. Using these models, we show how the ultrafast inverse Faraday effect induces a net magnetic moment in antiferromagnets and demonstrate that a crystal field environment and the exchange interaction play essential roles for laser-induced magnetization dynamics even during the action of a pump pulse. Using our approach, we show that light-induced precessions can start even during the action of the pump pulse with a duration several tens times shorter than the period of induced precessions and affect the position of magnetic vectors after the action of the pump pulse.




\end{abstract}
\pacs{75.78.Jp,75.40.Gb,78.20.Ls}
\maketitle

\section{Introduction}

Ultrafast optical control of a magnetic state of a medium is a rapidly developing field of research \cite{KirilyukRMP10,KimelNatRevMat19}. Laser-induced magnetization dynamics can take place on a subpicosecond time scale providing the possibility to overcome the time limit of several picoseconds for precessional magnetic switching. Therefore, laser manipulation techniques are extremely promising for the development of data operation devices which are several orders of magnitude faster than that available now. However, despite the importance of subpicosecond laser-induced magnetization effects for technological applications, their origin is poorly understood.


The inverse Faraday effect (IFE) that leads to nonthermal induction of magnetization by circularly-polarized laser pulses \cite{Pitaevskii,Pershan} is one of magneto-optical effects, which can take place on a femtosecond time scale. The ultrafast inverse Faraday effect (UIFE) is particularly important for potential applications in magnetic recording and magneto-optical devices, since it enables nonthermal coherent control of magnetization dynamics at a subpicosecond time scale. It avoids problems caused by material heating, which limits a repetition frequency due to a required cooling time and a recording density due to heat diffusion. Therefore, the demonstration of the UIFE, in which magnetic oscillations in canted antiferromagnet DyFeO$_3$ were induced by circularly polarized ultrashort laser pulses \cite{KimelNature05}, motivated intensive theoretical \cite{WoodfordPRB09, PopovaPRB11, PopovaPRB12, MyThesis, QaiumzadehPRB13, BoseJournalOptics14, BattiatoPRB14, MondalPRB15, ZhangEPL16, ZhangMPLB16, MondalJPCM17, MurakamiJPCM17,PopovPRB21} and experimental \cite{ReidPRB10, MikhaylovskiyPRB12, MikhaylovskiyPRB15, ChoiNature17, KozhaevSciRep18} studies of this process. A significant progress in development of techniques of ultrafast spin control using femtosecond laser pulses based on the UIFE was demonstrated in recent years \cite{SatohPRL10, SatohNature12, SatohNature14, ParchenkoAPL16, ChoiNature17, SatohNature17,YoshimineEPL17,ChernovPhotonRes18, WangOptLett18, ImPRB19, NemecNatPhys18}. It has recently became possible to gain atomic- and spin-selective real-time insight into light-induced magnetization dynamics during the excitation by a pump pulse employing attosecond extreme ultraviolet pulses \cite{SiegristNature19, HofherrSciAdv20, WillemsNatComm20}. A big effort is being performed at Free-Electron Laser Facilities such as Linear Coherent Light Source LCLS  to enable attosecond x-ray imaging experiments \cite{DurisNatPhot20}. Such attosecond x-ray pulses can provide a detailed insight into real-time laser-driven electron and spin dynamics \cite{Popova-GorelovaAppSci18,Popova-GorelovaArxiv20Short}. These advances give rise to a demand to better understand light-induced magnetization dynamics at ultrashort time scales.





Equations of motion for a magnetic moment are a standard tool to describe magnetization dynamics.  Usually, it is straightforward to derive magnetic equations of motion after the action of a pump pulse taking the new laser-induced magnetic state as an initial condition. Landau-Lifschitz-Bloch equation is usually applied for a macroscopic description of magnetization dynamics \cite{KirilyukRMP10,AtxitiaPRB11,NievesProceedings15, MorenoPRB17}, which can be derived from a microscopic description within the Heisenberg representation. However, the ability to fully control magnetization dynamics in a material requires not only a description of magnetization dynamics after the excitation, but also an ability to calculate evolution of a magnetic moment due to the action of a pump pulse depending on its parameters and material properties. An inclusion of a coupling between a magnetic system and a pump pulse into magnetic equations of motion is necessary for this goal. 

The phenomenological description of the IFE \cite{Pitaevskii,Pershan} suggests that the action of a circularly polarized laser pulse should be considered as an effective magnetic field, which is proportional to the pulse intensity. However, it was shown by us theoretically \cite{PopovaPRB11, PopovaPRB12,MyThesis} and demonstrated in several experimental works \cite{ReidPRB10, MikhaylovskiyPRB12} that this description of the IFE is not applicable at subpicosecond time scales, since it was developed for laser pulses with duration much longer than system relaxation times. Thus, a proper description of this process at subpicosecond time scales is necessary for further achievements in the field of ultrafast magnetization dynamics control. 

A sudden approximation can be used as an alternative to the phenomenological description \cite{PerroniPRB06, GridnevPRB08, IidaPRB11, KalashnikovaPRB08}. Within this approach, the effect of a laser pulse is considered as an action of an ultrashort magnetic pulse, amplitude of which is proportional to light intensity multiplied by the Verdet constant. However, there are two disadvantages of this approach. First, if the action of a laser pulse is substituted by a magnetic field, then a new magnetic state after the excitation cannot be predicted  accurately and, hence, this approach cannot be applied for a development of a mechanism to control magnetization dynamics. The second problem is that the condition that a laser pulse duration is much shorter than a system's oscillation period is not valid in many cases. For instance, it is not applicable to the description of experiments demonstrating light-induced terahertz precessions\cite{ReidPRL10, SatohPRL10,  ParchenkoAPL16, SatohNature14}, when a magnetic oscillation with a period of one or several picoseconds is induced by a laser pulse of several hundreds of femtoseconds duration. 

In this article, we derive an effective magnetic operator expressed in terms of total angular momentum operators that accurately describes magnetization dynamics at a subpicosecond time scale during the action of an ultrashort laser pulse via the UIFE. The time-dependent functions entering the operator depend on parameters of a laser pulse and the coupling of the electric field of a laser pulse to the electronic system of a material. They are nonzero during the action of a laser pulse and zero afterwards. The action of the effective magnetic operator is separated from that of other magnetic operators describing fields acting on a magnetic system apart from light, such as an external magnetic field, exchange interaction {\it etc}. Thus, after the action of the laser pulse, the action of the effective magnetic operator turns off and magnetization dynamics persist due to a deviation of a magnetic vector from its ground state. 

The equations of motion of magnetic vectors are obtained from the commutators of total angular momentum operators with the effective magnetic operator and other magnetic operators describing fields acting on a magnetic system. This way, we calculate magnetization dynamics during and after the action of light within the Heisenberg representation. We show that magnetization dynamics obtained in the Heisenberg picture are equivalent to that obtained with the time-dependent Schr\"odinger equation, which describes the perturbation of an electronic system by the electric field of a laser pulse. The first advantage of the Heisenberg representation over the Schr\"odinger representation is that it shows how the optical process affects components of the total angular momentum individually. And, most importantly, the Heisenberg representation provides a link for the derivation of the description of the UIFE at a nanoscale \cite{EvansJPhConMatt14}.

In our previous studies, we described the effect of the spin-orbit coupling on the IFE and how light-induced electronic transitions lead to the change of a spin state \cite{PopovaPRB11, PopovaPRB12}. In this article, we also consider the effects of the Zeeman, exchange and crystal-field interactions, and include their corresponding effective magnetic Hamiltonians into the derived equations of motion. We find that the exchange interaction and, especially, the crystal-field interaction can considerably influence the dynamics of magnetic vectors even during the action of the pump pulse. The approach derived in this article is able to describe the interplay between the deviation of magnetic vectors due to the action of the pump pulse and light-induced magnetic precessions launched due to this deviation. As we will show, this effect can be substantial even if the pump-pulse duration is several tens of times shorter than the period of precessions.


The article is organized as follows. We derive the effective magnetic operator operator and the equations of motion of magnetic moment due to the action of the UIFE in the Section I. We apply this formalism to describe the dynamics of a single spin system in an external magnetic field during and after the excitation by a circularly polarized laser pulse in Section II. We show that an accurate calculation of the time evolution of a magnetic vector during the action of the pump pulse is necessary even if the period of the induced oscillations is several tens of times longer than the laser pulse duration.  We derive equations of motion for magnetic vectors of a system consisting of two antiferromagnetically coupled sublattices during and after the action of an ultrashort laser pulse in Section III. With the help of the equations of motion, we describe the mechanism of the generation of a non-zero magnetic moment and its precessions in compensated antiferromagnets due to the UIFE. It is shown that a crystal field environment and exchange and crystal field coupling have a strong effect on laser-induced magnetization dynamics even during the action of a pump pulse.


\section{Derivation of the effective magnetic operator}

The UIFE leads to a nonthermal change of a magnetic state of a system by a circularly-polarized laser pulse via the stimulated Raman scattering process \cite{Pershan,PopovaPRB11, PopovaPRB12,MyThesis}. Thereby, the laser pulse excites electron transitions in the system in such a way that the initial and the final states belong to the same ground state manifold, but are energetically separated by internal or external magnetic interactions \cite{ShenBook, ShenPhRev66}. The magnetic state of this electronic system is described by a spinor $\Psi_g$ with components, which are projections of the wave function of the system on the eigenstates of the total angular momentum. Since the UIFE leads to the change of a magnetic state of the electronic system, it must be possible to introduce an effective magnetic operator $\Hs$, which describes this effect. $\Hs$ must act on the spinor $\Psi_g$ and fulfil the time-dependent Schr\"odinger equation  
\begin{equation}
i\Psi'_g=[\hat{\mathcal H}_m+\Hs]\Psi_g \label{SchrodHeisen}.
\end{equation}
Here and throughout this article, we use the atomic units. $\mathcal H_m$ is the magnetic Hamiltonian, which includes all external and internal magnetic interactions acting on the total angular momentum apart from $\Hs$. If the operator $\Hs$ is known, one can determine the equations of motion of projections $J_x$, $J_y$ and $J_z$ of the total angular momentum due to the action of the IFE using the Heisenberg representation as
\begin{align}
i J_{\alpha}' = \left\l\Psi_g\left| \left[\hat J_{\alpha},\hat{\mathcal H}_m+\Hs\right]\right|\Psi_g\right\r\label{Heisen},
\end{align}
where $J_{\alpha} = i \left\l\Psi_g\left|\hat J_{\alpha}\right|\Psi_g\right\r$, $\alpha$ stays for $x$, $y$ and $z$. 



In order to derive $\Hs$, we take into account that the time evolution of spinor $\Psi_g$ obtained with the time-dependent Schr\"odinger equation in Eq.~(\ref{SchrodHeisen}) should be equivalent to the one derived from the solution of time-dependent Schr\"odinger equation for the wave function $\widetilde\Psi(t)$ of the electronic system
\begin{equation}
i\widetilde\Psi'(t)=(\hat{\mathcal H}_0-\mathbf d\cdot \mathbf E)\widetilde\Psi(t)\label{SchrEq}.
\end{equation}
$\hat{\mathcal H}_0 = \sum_i\mathbf p_i^2+\hat{\mathcal V}$, where $\mathbf p_i$ is the momentum of an electron, the summation is over all electrons in the system. $\hat{\mathcal V}$ includes the kinetic energy of nuclei, the interaction energy between electrons and nuclei, mutual Coulomb energy of the electrons and nuclei, and all internal and external magnetic interactions \cite{PopovaPRB11, PopovaPRB12, MyThesis}.  $\mathbf d$ is the dipole moment of the system. $\mathbf E$ is the electric field of a laser pulse with a frequency $\omega_0$ 
\begin{equation}
\mathbf E=\mathbf n\mathcal E \, p(t/T-\mathbf r/(cT))\sin(\omega_0t)\label{ElField}.
\end{equation}
The electronic system is assumed to have the spatial extend much smaller than the wavelength $\lambda_0=c/\omega_0$ is considered. Also it is assumed that $\lambda_0\ll cT$, thus the pulse spatial dependence is ignored. $\mathcal E$ is the amplitude of the electric field, $p(t/T)$ describes the time-dependence of the amplitude of the electric field. $\mathbf n$ is a unit vector perpendicular to the light propagation direction. Throughout this article, we consider the action of a left-circularly polarized laser pulse propagating in the $z$ direction, which corresponds to $\mathbf n =(\mathbf n_x+i\mathbf n_y)/\sqrt 2$.

The solution of the Eq.~(\ref{SchrEq}) can be represented by an expansion in functions $\widetilde\Psi_n$ of order $(1/c)^n$
\begin{align}
\widetilde\Psi(t)=\U&(\widetilde\Psi_0+\widetilde\Psi_1(t)+\widetilde\Psi_2(t)+\hdots)\nonumber\\
=\U&\left(\widetilde\Psi_0-i\int_{-\infty}^{t}dt'\,\U^{-1}\hat V\U\widetilde\Psi_0 \right.\label{ExpansionInt}\\
&-\left.\int_{-\infty}^tdt'\,\U^{-1}\hat V\U\int_{-\infty}^{t'}dt''\,\U^{-1}\hat V\U\widetilde\Psi_0+\hdots\right),\nonumber
\end{align}
where $\widetilde\Psi_0=\widetilde\Psi(0)$, $\hat V=- \mathbf A\cdot\sum_{i}\mathbf p_{i}/c$, $\mathbf A$ is a vector potential, which is related to the electric field by $\mathbf E=-\dot{\mathbf A}/c$. $\U$ is the time evolution operator, which fulfils the equation $i\,\U'=\hat{\mathcal H}_0\, \U$. 
The IFE experiments are typically done at frequencies corresponding to a material transparency region, where the absorption is very weak. The intermediate states can be considered as virtually excited, and the contribution of the first order wave function is not taken into account. The stimulated Raman scattering is described by the second order wave function $\widetilde\Psi_2(t)$. Thus, the normalized spinor $\Psi_{\text{g}}(t)$ is obtained by
\begin{equation}
\Psi_{\text{g}}(t)=\frac{\U\bigl(\Psi_0+\Psi_2(t)\bigr)}{\|\Psi_0+\Psi_2(t)\|}\label{Psig},
\end{equation}
where $\Psi_0$ and $\Psi_2(t)$ are the spinors associated with the wave functions $\widetilde\Psi_0$ and $\widetilde\Psi_2(t)$. 

The spinor $\Psi_0$, which describes the magnetic state of the electronic system before the action of light, can be represented as
\begin{equation}
\Psi_0=
\begin{pmatrix}
P_{01}\\P_{02}\\ \vdots\\ P_{0n}
\end{pmatrix}=P_{01}\Bigl|J,J_z=J\Bigr\r+P_{02}\Bigl|J,J-1\Bigr\r+\dotsc,
\end{equation}
where $J$ is the total angular magnetic momentum, $n=2J+1$ and $P_{0k}$ is the projection of $\Psi_0$ on the state $|J,J_z = J+1-k\Bigr\r$. $\sum_k|P_{0k}|^2=1$ is the normalization condition.

$\Psi_2(t)$ is a spinor with $n$ time-dependent components. Each component of this spinor is given by a transition amplitude of stimulated Raman scattering to a final state with a corresponding projection of $J_z$ (see Ref.~\onlinecite{PopovaPRB11, PopovaPRB12,MyThesis} for details). According to the selection rules for the stimulated Raman scattering process induced by a circularly polarized laser pulse propagating in the $z$ direction, a transition from a state $|J,J_z\r$ back only to the same state is allowed. This means that the $k$-th component of $\Psi_2(t)$ is a transition amplitude $T_k$ describing transition from an initial state $|J,J+1-k\r$ to a final state $|J,J+1-k\r$. Due to the presence of the spin-orbit coupling, transition amplitudes $T_k$ are different for each $k$. Thus, the spinor $\Psi_2(t)$ is not proportional to $\Psi_0$ and describes a magnetic state, which is different from the initial one.

We introduce time-dependent factors $\mathcal A_k(t)$ such that the $k$-th element of the spinor $\Psi_0+\Psi_2(t)$ is equal to $\mathcal A_k(t)P_{0k}$ and satisfy $\mathcal A_k(0)=1$.  The factors $\mathcal A_k(t)$ are determined by Eq.~(\ref{ExpansionInt}) and depend on the electric field of the laser pulse, selection rules and energies of system's ground and excited states, but does not depend on the initial state of the system. Applying Eq.~(\ref{Psig}), $\Psi_g(t)$ can be represented as
\begin{equation}
\Psi_g=
\frac{\U}{\norma(t)}
\begin{pmatrix}
\mathcal A_1(t)P_{01}\\\vdots\\\mathcal A_k(t)P_{0k}\\\vdots
\end{pmatrix}=
\begin{pmatrix}
P_{1}(t)\\\vdots\\P_{k}(t)\\\vdots
\end{pmatrix}\label{Adef},
\end{equation}
where $\mathcal N(t)=\|\Psi_0+\Psi_2(t)\|$ is the normalization factor. 


We derived the operator $\Hs$ providing $\Psi_g$ from Eq.~(\ref{SchrodHeisen}), which is equivalent to the solution of Eq.~(\ref{Psig}) (see Appendix \ref{MomentumOperatorAppendix} and Ref.~\onlinecite{MyThesis} for the details). We obtain that
\begin{equation}
\Hs=\begin{pmatrix}
\ddots&&&&\\
&-\gamma_a&\cdots&i P_a P_b^*\left(\nu_a-\nu_b\right)&\cdots\\
&\vdots&\ddots&&\\
&i P^*_a P_b\left(\nu_b-\nu_a\right)&&&\\
&\vdots&&&
\end{pmatrix},
\end{equation}
where $a=1\dotsc n$ is the row number and $b=1\dotsc n$ is the column number. Namely, the diagonal elements of the operator are $[\Hs]_{aa} = -\gamma_a(t)$, and the off-diagonal elements are $[\Hs]_{ab} = i P_a(t) P_b(t)^*\left(\nu_a(t)-\nu_b(t)\right)$. $P_a(t)$ and $P_b(t)$ are the components of $\Psi_g$, and
\begin{align}
&\nu_a(t)=\Re\left(Y_a(t)\right),\nonumber\\
&\gamma_a(t)=\Im\left(Y_a(t)\right),\label{FunDef}\\
&Y_a(t)={[\U\mathcal A']_a}/{[\U\mathcal A]_a}\nonumber,
\end{align}
where $\mathcal A$ is a spinor with elements $\mathcal A_a(t)$ and $\mathcal A'$ is a spinor with elements $\mathcal A'_a(t)$. $[\U\mathcal A']_a$ and $[\U\mathcal A]_a$ are the $a$-th components of the spinors $\U\mathcal A'$ and $\U\mathcal A$, correspondingly. 

One can express the operator $\Hs$ via operators $\hat N_{ab+}$ and $\hat N_{ab-}$ and their expectation values, where the elements of these operators are
\begin{eqnarray}
&&(\hat N_{ab+})_{ab}=(\hat N_{ab+})_{ba}=1,\text{ where }b\ge a,\nonumber\\
&&(\hat N_{ab-})_{ba}=(\hat N_{ab-})^*_{ab}=i,\text{ where }b>a,\nonumber\\
&&\hat N_a=\hat N_{aa+},\label{NkCh4}\\
&&(\hat N_{a})_{aa}=1,\nonumber\\
&&(\hat N_{ab\pm})_{mn}=0,\text{ if }m\neq a,\,m\neq b,\, n\neq a,\,n\neq b\text{ or } b<a.\nonumber
\end{eqnarray}
For example, if $J=3/2$, these operators are
\begin{align}
&\hat N_{12+}=
\begin{pmatrix}
0&1&0&0\\
1&0&0&0\\
0&0&0&0\\
0&0&0&0
\end{pmatrix},\nonumber\\
&\hat N_{12-}=
\begin{pmatrix}
0&-i&0&0\\
i&0&0&0\\
0&0&0&0\\
0&0&0&0
\end{pmatrix},\label{Nab3/2}\\
&\hat N_1=
\begin{pmatrix}
1&0&0&0\\
0&0&0&0\\
0&0&0&0\\
0&0&0&0
\end{pmatrix}.
\nonumber
\end{align}
Applying that $P_a P_b^* = \l\Psi_g|\hat N_{ab+}-i\hat N_{ab-}|\Psi_g\r/2$ and $P_a^* P_b = \l\Psi_g|\hat N_{ab+}+i\hat N_{ab-}|\Psi_g\r/2$, the effective magnetic operator $\Hs$ can be represented as
\begin{align}
\Hs=&-\sum_a^n \gamma_a \hat N_a\label{MomOperator}\\
&+\frac12\sum_{a,b}^n\left(\nu_a-\nu_b\right)\left( \la \hat N_{ab-} \ra \hat N_{ab+}- \la \hat N_{ab+} \ra\hat N_{ab-}\right)\nonumber.
\end{align}
This representation allows deriving a convenient system of differential equations for the variables $\la \hat N_{ab\pm}\ra$, which are connected to the expectation values of the momentum operators $\hat J_{x}$, $\hat J_y$ and $\hat J_z$ as will be shown below.

The equations of motion for the expectation values of $\hat N_{ab\pm}$ operators are given by [see Eq.~(\ref{Heisen})]
\begin{align}
\la \hat N_{ab\pm}\ra'&=-i\lf\l \lf[\hat N_{ab\pm},\Hs+\hat{\mathcal H}_m\rt]\rt\r\label{EqMotGeneral0}
\end{align} 
The commutators $-i\lf\l \lf[\hat N_{ab\pm},\Hs\rt]\rt\r$ are derived in Appendix \ref{MomentumOperatorAppendix} and result in
\begin{align}
\la \hat N_{ab\pm}\ra' = &\lf(-2\sum_k\nu_k \la \hat N_k\ra+\nu_a+\nu_b\rt)\la \hat N_{ab\pm}\ra\nonumber\\
&\pm\lf(\gamma_a-\gamma_b\rt) \la \hat N_{ab\mp}\ra -i\lf\l \lf[\hat N_{ab\pm},\H\rt]\rt\r\label{EqMotionGeneral}.
\end{align}

The matrices representing total angular momentum operators $\hat J_x$, $\hat J_y$ and $\hat J_z$ can be expressed by linear combinations of $\hat N_{ab\pm}$ operators. For example, $\hat J_x = \sqrt3\hat N_{12+}/2+\hat N_{23+}+\sqrt3\hat N_{34+}/2$, $\hat J_y = \sqrt3\hat N_{12-}/2+\hat N_{23-}+\sqrt3\hat N_{34-}/2$ and $\hat J_z = 3\hat N_{1}/2+\hat N_2/2-\hat N_3/2- 3\hat N_4/2$ for $J = 3/2$. This way, the time evolutions of $J_x$, $J_y$ and $J_z$ can be obtained from the equations of motion for $\la \hat N_{ab\pm}\ra$ as will be shown with the examples presented in the Sections \ref{SectionSpin1/2} and \ref{SectionAntiferro}.


The advantage of our approach derived in the Heisenberg representation is that one can include the dissipation processes using the density matrix formalism and the Redfield equation \cite{Blum2013density}
\begin{align}
i \rho' = [ \hat{\mathcal H}_m+\Hs, \rho] -i \hat \Gamma (\rho-\rho_{\text{eq}}).
\end{align}
Here, $\mathcal H$ is the total magnetic Hamiltonian, ${\mathcal H}_m$ is the magnetic Hamiltonian in the absence of the light field and $\Hs$ is the effective magnetic Hamiltonian due to the action of the pump pulse that we derived in our manuscript. $\rho=|\Psi_g\ra\la\Psi_g|$ is the time-dependent density matrix in the Hilbert space of the magnetic system, $\rho_{\text{eq}}$ is the density matrix at the equilibrium. $\hat\Gamma$ is the relaxation superoperator. For example, it can be diagonal with the matrix elements being relaxation rates. The time evolution of the magnetic vectors is then derived using the trace of $\rho$ acting on corresponding magnetic operators, e.g.~$\operatorname{Tr}[\rho J_x]$. We do not consider the dissipation in the processes described in the following Sections, since its treatment is beyond our study.

\section{Single spin dynamics due an external magnetic field and UIFE}
\label{SectionSpin1/2}
\subsection{Equations of motion}
\label{SubsecSingleSpinEqs}

\begin{figure}[t]
\centering
\includegraphics[width=0.45\textwidth]{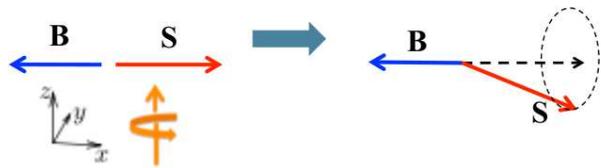} 
\caption{Spin 1/2 in an external magnetic field}
\label{Question}
\end{figure}

\begin{figure}[t]
\centering
\includegraphics[width=0.3\textwidth]{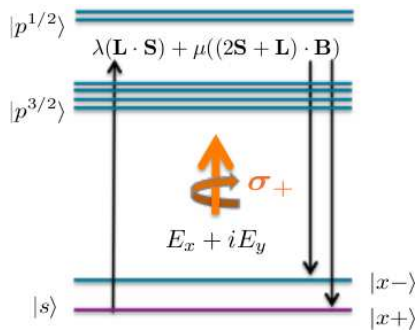} 
\caption{The energy level scheme of the single-spin system.}
\label{LevelsCh5}
\end{figure}

In this Section, we describe single-spin dynamics due to a joint action of the UIFE and an external magnetic field $\mathbf B$. We assume that an external magnetic field $\mathbf B$ is aligned in the $-x$ direction and the gyromagnetic ratio is equal to 1 (see Fig.~\ref{Question}). The ground state is split into two states: spin parallel, $|x-\rangle$, and antiparallel, $|x+\rangle$, to the magnetic field with the corresponding energies $\epsilon_{x-}=\epsilon_{1s}+B/2$ and $\epsilon_{x+}=\epsilon_{1s}-B/2$, where $\epsilon_{1s}$ is the ground state energy in the absence of the magnetic field. The system is initially in the lowest energy state with the spin aligned in the $+x$ direction, which is described by the spinor $\Psi_0=\left(\begin{smallmatrix}1/\sqrt2\\1/\sqrt2\end{smallmatrix}\right)$. A left-circularly polarized ultrashort laser pulse propagating in the $z$ direction triggers Raman transitions via a virtual excited state $2p$, which is split due to the spin orbit coupling $\lambda(\mathbf L\cdot \mathbf S)$ and Zeeman interaction $\mu((2\mathbf S+L)\cdot \mathbf B)$ into six states (see Fig.~\ref{LevelsCh5} and Appendix \ref{AppMagField}). Here, $\mu$ is the gyromagnetic ratio, $\lambda$ is the spin-orbit coupling constant and $\mathbf L$ is orbital momentum. We assume $\mu=-1/2$ and $\lambda = 20$ meV. The UIFE leads to a change of a spin state, thereby the spin deviates from its initial position and starts precessing due to the magnetic field.

The total magnetic Hamiltonian $\hat{\mathcal H}_m^{\text{tot}}$, which determined spin dynamics, is the sum of the effective magnetic operator $\Hs$ and Hamiltonian describing the Zeeman interaction with the external magnetic field
\begin{equation}
\hat{\mathcal H}_m^{\text{tot}}=\Hs-B_x\hat S_x.
\end{equation} 
In the case of $S=1/2$, the spin operators are linear combinations of the $\hat N_{ab\pm}$ operators: $\hat S_x=\hat N_{12+}/2$, $\hat S_y=\hat N_{12-}/2$, $\hat S_z=(\hat N_1-\hat N_2)/2$, $\hat S^2=3(\hat N_1+\hat N_2)/2$. Thus, the operator $\Hs$ can be represented as
\begin{equation}
\Hs=f(t)\left( S_y\hat S_x- S_x\hat S_y\right)+g(t)\hat S_z+h(t)\hat S^2, \label{UfIFE_Spin1/2}
\end{equation} 
where $f(t)=2\Re(Y_2(t)-Y_1(t))$, $g(t)=\Im(Y_2(t)-Y_1(t))$ and $h(t)=-\frac23\Im(Y_2(t)+Y_1(t))$, $S_{\alpha} = \la \hat S_{\alpha}\ra$ for $\alpha=x,y,z$. The function $Y_1(t)$ is determined by the transition amplitude of the Raman transitions from the state $|1s,S_z=+\frac12\rangle$ back to the state $|1s,S_z=+\frac12\rangle$ and $Y_2(t)$ is determined by the transition amplitude the Raman transitions from the state $|1s,S_z=-\frac12\rangle$ to the state $|1s,S_z=-\frac12\rangle$ [see Eq.~(\ref{FunDef})].

The third term of $\Hs$ in Eq.~(\ref{UfIFE_Spin1/2}) determined by function $h(t)$ is proportional to the identity matrix and does not influence the spin. The action of the first and the second terms of $\Hs$, which are determined by the functions $f(t)$ and $g(t)$, lead to spin rotation. Both functions $f(t)$ and $g(t)$ are given by the difference between $Y_1(t)$ and $Y_2(t)$. Thus, the larger the difference between the transition amplitudes for spin-up and spin-down states is, the larger functions $f(t)$ and $g(t)$ are, and the more effective the spin is rotated by light [see Eq.~(\ref{FunDef})]. If transition amplitudes for different spin components are equal, then $Y_1(t)=Y_2(t)$, $\Hs$ is proportional to the identity matrix, and no rotation of spin by light is possible.

The equations of motion for the spin vector can be derived either from the general equation (\ref{EqMotionGeneral}) or from the relation $S_{\alpha}'=-i\langle[\hat S_{\alpha},\hat{\mathcal H}_m^{\text{tot}}]\rangle$ resulting in
\begin{align}
&S_x'=-f(t)S_xS_z-g(t)S_y,\nonumber\\
&S_y'=-f(t)S_yS_z+g(t)S_x+B S_z,\label{SpinEqMot}\\
&S_z'=-f(t)(S_x^2+S_y^2)-B S_y.\nonumber
\end{align}
It follows from these equations that the first term of the effective operator $\Hs$ determined by the function $f(t)$ results in a quadratic effect on a spin and the second term determined by $g(t)$ - in a linear one. Both terms describe rotation of a spin around some axis with a time-dependent frequency. The first term describes nonlinear rotation of spin around an axis, which is perpendicular to the light propagation direction and to the initial direction of spin. The second part determined by $g(t)$ describes rotation of the spin around the $z$ axis. This is in agreement to a conclusion that a spin is rotated only around the light propagation direction via the UIFE, if a laser pulse has a frequency far from a resonance\cite{PopovaPRB12}. In this case, both functions $f(t)$ and $g(t)$ are relatively small since the transition amplitudes are low and, consequently, the quadratic effect is stronger suppressed than the linear one.



The functions $f(t)$ and $g(t)$ can be calculated as follows. According to Eq.~(\ref{FunDef}), the functions $Y_{1}(t)$ and $Y_{2}(t)$ depend on the spinor $\mathcal A(t)$ and the time evolution operator $\U$. The time evolution operator, which fulfils the relation $i\,\U'=\hat H_m\,\U$, is $\U=e^{-iB \hat S_xt}$. The spinor $\mathcal A(t) = \left(\begin{smallmatrix}1+\psi_{2\uparrow}(t)\\1+\psi_{2\downarrow}(t)\end{smallmatrix}\right)$ is determined by the transition amplitudes of the Raman scattering process for the spin-up state and spin-down state, $\psi_{2\uparrow}(t)$ and $\psi_{2\downarrow}(t)$, correspondingly. According to Eq.~(\ref{ExpansionInt}), 
\begin{align}
\mathcal A(t) = \begin{pmatrix}1\\1\end{pmatrix}+
\left(\frac{\mathcal E}{\omega_0}\right)^2\int_{-\infty}^tdt'\,\U(t')
\begin{pmatrix}
\sum_j |d_{\uparrow j}|^2 G_{j}(t')\\
\sum_j |d_{\downarrow j}|^2 G_{j}(t')
\end{pmatrix},\label{AforSpin1/2}
\end{align}
where
\begin{align}
&G_j(t')=e^{-i\Delta\omega_{0j}t'}F(t')\int_{-\infty}^{t'}dt''e^{i\Delta\omega_{0j}t''}e^{-iBt''/2}F(t''),\nonumber\\
&F(t)=p(t/T)\cos(\omega_0t)\label{Ft}.
\end{align}
Here, the summation is over intermediate states $j$. $\Delta\omega_{0j}=\epsilon_{2p,j}-\epsilon_{1s}$,  $\epsilon_{2p,j}$ are the energies of the intermediate states. $d_{\uparrow j}$ and $d_{\downarrow j}$ are the dipole matrix elements of the transitions from the states $|1s,S_z=1/2\rangle$ and $|1s,S_z=-1/2\rangle$ to the state $j$. The dipole matrix elements are calculated in Appendix \ref{AppMagField}. It is also shown in Appendix \ref{AppMagField} that a spin state can be changed via the Raman scattering process only if the spin-orbit coupling is nonzero even if an external magnetic field is nonzero.

As shown in the next Subsection, the functions $f(t)$ and $g(t)$ are nonzero during the action of a laser pulse and smoothly become zero at time $\tau_{\text p}$, when the excitation is finished. Thus, during the action of a laser pulse, Eq.~(\ref{SpinEqMot}) describes the spin motion due to both laser excitation and the external magnetic field. The terms in Eq.~(\ref{SpinEqMot}), which describe spin motion due to the coupling to the electromagnetic field, are smoothly turning off while the excitation is finishing, and spin moves only due to an external magnetic field after the action of a laser pulse. Thus, we have obtained a system of differential equations with terms separately describing spin motion due to the Zeeman interaction and due to the UIFE, which shows the interplay of the interactions leading to the rotation of spin.

\subsection{Time evolution of the spin vector}
\label{SectionTimeEvolSinhleSpin}

\begin{figure}[t]
\centering
\includegraphics{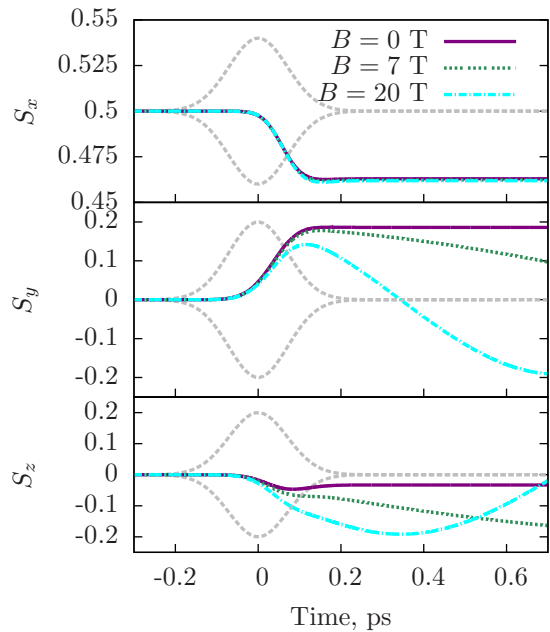} 
\caption{Time evolution of the spin-vector components due to the excitation at different applied magnetic fields in the $x$ direction. The gray line represents the time evolution of the electric field amplitude. Laser pulse duration is 117 fs.}
\label{Sxyz}
\end{figure}

We calculate and compare laser-induced spin dynamics in the presence of an external magnetic field with a magnitude of $7$ T and of $20$ T. Although the chosen magnetic field magnitudes are rather high, they are reasonable for a goal of comparison to the experiments studying the UIFE in the presence of an applied external magnetic field. Although external magnetic fields with magnitudes of up to 0.5 T are usually applied in experiments, materials under consideration obtain gyromagnetic factor about ten times higher than that of a single electron spin (see {\it e.\! g.\!} Refs.~\onlinecite{KalashnikovaPRB08, ReidPRB10, HansteenPRB06,JinAPL10}). Thus, the chosen magnitudes of the magnetic fields result in Larmor precession frequencies comparable to the ones in experiments.

We consider the action of a left-circularly polarized Gaussian shaped laser pulse with $p(t/T)=\exp(-t^2/T^2)/\sqrt{\pi^3}$ and fluence $E_{\text{fl}}\approx$ 2 mJ/cm$^2$. We assume $T=100$ fs, which corresponds to the pulse duration of 117 fs at FWHM of the intensity and the bandwidth of 15 meV. The laser pulse central frequency $\omega_0$ is equal to $(\epsilon_{2p}-\epsilon_{1s})$, where $\epsilon_{2p}$ is the energy of the unsplit $2p$ state, i.e.~the energy of the state in the absence of the Zeeman and spin-orbit interactions. The Larmor precession periods of the magnetic fields of 7 T and 20 T are approximately 5 ps and 1.7 ps, correspondingly (see Table \ref{TableRatio}). The ratios of the induced precession periods to the pulse duration are 40 and 15, respectively. 

\begin{table}[t]
\begin{center}
\begin{tabular}{|c|c|c|}
\hline
&&\\
$B$ & $T_L$ & $T_L/T_{\text{dr}}$\\
\hline
7 T & 5 ps & 40\\
\hline
20 T & 1.7 ps & 15\\
\hline
\end{tabular}
\end{center}
\caption{Larmor periods $T_L$ and their ratios to the laser pulse duration, $T_L/T_{\text{dr}}$, for the chosen magnetic fields.}
\label{TableRatio}
\end{table}

\begin{figure}
\centering             
\includegraphics{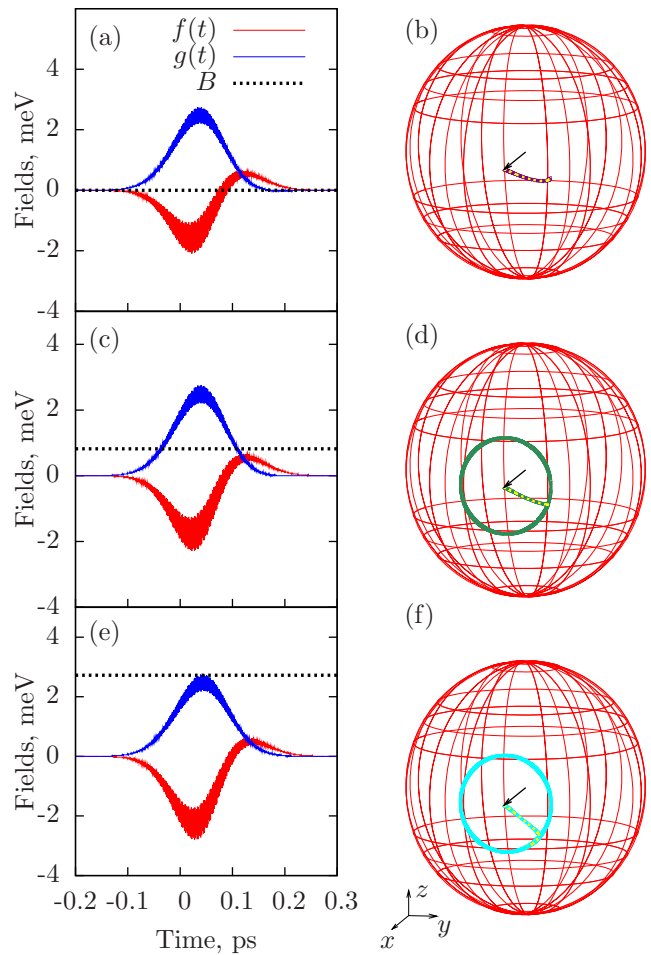}
\caption{Left column: fields (in energy units) acting on the spin during the excitation. Right column: the corresponding time evolution of the spin vector on the Bloch sphere. The black arrow shows the initial alignment of spin. The dotted lines on the left plots show the dynamics of the spin during the excitation, {\it i.\ e.\!} at $t<200$ fs, the continuous lines show the dynamics of the spin after the excitation at $t>200$ fs. The external magnetic fields are (a), (b) $B=0$. (c), (d) $B=7$~T. (e), (f) $B=20$ T.}
\label{Forces_and_3d}
\end{figure}

The time evolutions of the spin-vector components $S_x$, $S_y$ and $S_z$ are shown on Fig.~(\ref{Sxyz}) for $B=0$, 7 T and 20 T. $S_x$ component is influenced only due to the interaction with light and its dynamics does not depend on the magnetic field [see Eq.~(\ref{SpinEqMot})]. But comparing the time evolutions of the spin components $S_y$ and $S_z$ at different magnetic fields, one can see that they noticeably depend on the magnitude of the magnetic field even during the interaction with light. The values of $S_y$ and $S_z$ at time $\tau_{\text{p}}=200$ fs, when the excitation is negligible, are very strongly affected by the magnetic field. This demonstrates that Larmor precession has a considerable impact on laser-induced spin dynamics already during the action of the laser pulse.

A sudden approximation suggests that the laser excitation is immediate and thus the spin oscillation in an external magnetic field during the excitation can be ignored. Thus, one may assume that spin Larmor precession starts after the excitation, namely, at time $\tau_{\text{p}}=200$ fs. The spin state at $\tau_{\text{p}}$ calculated without an applied magnetic field would be taken as the initial condition for the precession. Applying this assumption and comparing it to the results obtained with Eq.~(\ref{SpinEqMot}), we obtain that the phase disagreement between spin dynamics calculated with and without this assumption are 14$^{\circ}$ in the case of $B=7$ T (the corresponding Larmor frequency of 5 ps) and 47$^{\circ}$ in the case of $B=20$ T (the corresponding Larmor frequency of 1.7 ps). Thus, the sudden approximation does not work correctly even if the oscillation period is about fifty times larger than the pulse duration. This statement is confirmed by the observation of Satoh {\it et al.\!} in Ref.~\onlinecite{SatohPRL10} that models, which ignore the time-dependency of a laser pulse, are not sufficient to describe the initial state of a magnetic precession.

Fig.~\ref{Forces_and_3d} shows the 3D picture of spin-vector trajectories during and after the excitation. This Figure additionally demonstrates how an applied magnetic field influences the spin dynamics. When the magnetic field is zero, the spin moves during the action of the laser pulse almost always in the $xy$ plane. But the spin trajectories during the excitation in the presence of the external magnetic fields are rotated around the $z$ axis relatively to the trajectory at a zero magnetic field. Furthermore, the spin trajectory starts to follow that of the Larmor oscillation even during the action of the laser pulse at $B=20$ T corresponding to the Larmor frequency of 1.7 ps (see Fig.~\ref{Forces_and_3d}f).

Figure~\ref{Forces_and_3d} compares the fields acting on the spin due to the magnetic field, $B$, and the interaction with light, $f(t)$ and $g(t)$ [see Eq.~(\ref{SpinEqMot})]. We obtain that $f(t)$ and $g(t)$ during the excitation are of the same order of magnitude as that of the chosen magnetic fields. Although $f(t)$ and $g(t)$ depend on the magnetic field via their dependence on the operator $\U=e^{-i \hat S_xB}$ [see Eq.~(\ref{FunDef})], this dependence is negligible even at the magnetic field of $20$ T. Since the bandwidth of the laser pulse, which is equal to 15 meV in the considered case, is much larger than energy splittings induced by the Zeeman interaction, the transition amplitudes are almost not affected by the Zeeman interaction. Thus, the dependence of the functions $f(t)$ and $g(t)$, which are determined by the transition amplitudes, on the magnetic field is negligible.

We have shown that the action of an external magnetic field can strongly affect spin dynamics and lead to an accumulation of a phase of the Larmor precession even during the excitation by light. Our simple model demonstrates that an exact calculation of the time evolution of the magnetic moment during the action of a laser pulse is necessary even if the pulse duration is several tens of times shorter than the period of an induced magnetic precession.


\section{Dynamics of an antiferromagnet}
\label{SectionAntiferro}

In this Section, we apply the Heisenberg picture to describe magnetization dynamics of two types of antiferromagnets, an easy-plane and an easy-axis antiferromagnet. We demonstrate the mechanism of the excitation of magnetic precessions and induction of a magnetic moment by the UIFE. A study of antiferromagnet dynamics is especially interesting, since magnetic resonances of such a system would not be observed, if a circularly-polarized laser pulse could be treated as an effective magnetic field. If it were true, magnetic precessions would be possible only in a material with a net magnetic moment, so that the effective magnetic field can produce a torque to it, which would result in magnetization precession \cite{KimelReview07}. However, light-induced terahertz magnetic precessions in antiferromagnets have been observed \cite{SatohPRL10, SatohNature14}.

The phenomenological model of Ref.~\onlinecite{GalkinJETP08} indeed predicted the possibility of the UIFE in an antiferromagnet. However, there are several problems to apply this approach to the interpretation of the experiments demonstrating light-induced terahertz magnetic precessions in antiferromagnets. First, the model is based on an assumption that the duration of a laser pulse is much shorter than the period of an induced spin precession. This approximation is not applicable for this experiment, since the ratio of the induced precessions periods to the pulse duration is about ten. Second, the model considers the light excitation as an ultrashort magnetic pulse and does not provide the information about the dependence of the effect on laser-pulse and material parameters.

The method introduced by us, first, does not make any assumptions on a pulse duration and, thus, can be applied to describe subpicosecond magnetization dynamics. Second, the technique involves the analysis of material properties and thus provides details about the dependence of the effect on a material structure. 

We apply several assumptions to treat the antiferromagnetic systems. However, our technique is not restricted to these assumptions, and other magnetic models can be chosen depending on a considered magnetic system. The model, which we use, is an example that can be adjusted to a realistic magnetic system.


\subsection{Antiferromagnetic systems}

\label{AntiferroStatesCh6}

We treat the exchange interaction in the framework of the Weiss mean field theory \cite{BuschowBook}. According to this theory, the quantum fluctuations can be neglected, and the exchange interaction between any two atoms is considered as the Zeeman interaction of a spin of each atom with a magnetic field, which is the spin average of the other atom. This means that the Hamiltonian $\hat{\mathcal H}_{\text{ex}12}=\mathcal J_{\text{ex}0}\hat S_1\cdot\hat S_2$ is substituted by $\hat{\mathcal H}_{\text{ex}}=\mathcal J_{\text{ex}0}(\hat S_1\l S_2\r+\l S_1\r \hat S_2)$. With the assumption that the exchange interaction only with the next neighbor atoms is relevant, the Hamiltonian acting on an atom $i$ is expressed as
\begin{equation}
\hat{\mathcal H}_{\text{ex}(i)}=Z\mathcal J_{\text{ex}0}\l S_{\text{nn}}\r \hat S_i,
\end{equation}
where $Z$ is the number of the neighboring atoms and $\l S_{\text{nn}}\r$ is the average spin of a next neighbor atom. Applying that the magnetic moment of an atom in the presence of the spin-orbit coupling is proportional to $g_J\mathbf J=\mathbf L+2\mathbf S$, where $g_J$ is the Land\'e factor and $\mathbf J=\mathbf L+\mathbf S$, the exchange interaction acting on atom $i$ can be expressed as
\begin{equation}
\hat{\mathcal H}_{\text{ex}(i)}=Z\mathcal J_{\text{ex}0}(g_L-1)^2\l J_{\text{nn}}\r\hat J_i.
\end{equation}
The approximation is valid, when the fluctuations of the effective magnetic field $Z\l J_{\text{nn}}\r$ are small, which is true, when each spin has many nearest neighbors.

We consider a system consisting of two equal sublattices coupled antiferromagnetically ($\mathcal J_{\text{ex}0}>0$).  Every atom belonging to the sublattice 1 is surrounded by $Z$ atoms belonging to the sublattice 2 and vice versa. The exchange interaction acting on atoms belonging to the sublattices 1 and 2 can be written in the framework of the Weiss mean field theory as
\begin{align}
&\hat{\mathcal H}_{\text{ex}}^{(1)} =\J\lf(J_{x2}\hat J_{x1}+J_{y2}\hat J_{y1}+J_{z2}\hat J_{z1}\rt),\\
&\hat{\mathcal H}_{\text{ex}}^{(2)} =\J\lf(J_{x1}\hat J_{x2}+J_{y1}\hat J_{y2}+J_{z1}\hat J_{z2}\rt)\nonumber,
\end{align}
where $\J=Z\mathcal J_{\text{ex}0}(g_L-1)^2$.

We treat the crystal field environment in the framework of the crystal field theory. Two types of uniaxial crystal fields are considered: one with the symmetry along the pulse propagation direction, $z$, and one in the $x$ direction, perpendicular to the pulse propagation direction. The spin-orbit coupling is assumed to be much larger than the crystal field, and the crystal field Hamiltonians can be expressed via the total angular momentum operators. The crystal fields are described by Hamiltonians $\Hcrz = \Delta_z\left(3\hat J_{z1,z2}^2-\hat J_{1,2}^2\right)$ and $\Hcrx = \Delta_x\left(3\hat J_{x1,x2}^2-\hat J_{1,2}^2\right)$, correspondingly. This assumption and the treatment of the crystal field are applicable to rare-earth-based magnetic materials \cite{BuschowBook} that are often used in magneto-optical experiments.

The magnetic Hamiltonian of the antiferromagnetic system in the case of the two types of crystal fields are
\begin{align}
&\hat{\mathcal H}_{m} = \hat{\mathcal H}_{\text{ex}}^{(1)}+\hat{\mathcal H}_{\text{ex}}^{(2)}+\label{H0Antiferro}\\
&\quad+\Delta_{z(x)}\left(3\hat J_{z1(x1)}^2-\hat J_1^2\right)+\Delta_{z(x)}\left(3\hat J_{z2(x2)}^2-\hat J_2^2\right).\nonumber
\end{align}
The alignment of the magnetic vectors in the ground state are determined by the sign of the crystal field constants $\Delta_{z,x}$ \cite{BuschowBook}. We chose the sign of both constants, $\Delta_z$ and $\Delta_x$ in such a way that in both cases the alignment of the magnetic vectors perpendicular to the pulse propagation direction, $z$, is energetically favorable. Thus, we choose $\Delta_z>0$, which results in the $xy$ plane being the easy plane (the $z$ axis becomes energetically unfavorable), and $\Delta_x<0$, which results in the $x$ being the easy axis.


Thus, in the ground-state, the magnetic vectors are aligned along the $x$ direction in the case of the crystal field $\Hcrx$, and are aligned along some direction in the $xy$-plane, which we define as the $x$ axis, in the case of the crystal field $\Hcrz$. Therefore, in both cases, the energy of the system is the lowest, when the absolute values of $J_{x1}$ and $J_{x2}$ are the largest, but the vectors $\mathbf M_{1}$ and $\mathbf M_{2}$ are antiparallel. We assume that the ground state is characterized by the term $J=3/2$, therefore, the spinors in both cases have initially the form  
\begin{eqnarray}
&\Psi^{(1)}_{0}=
\begin{pmatrix}
c\\d\\ d\\ c
\end{pmatrix},\ 
\Psi^{(2)}_{0}=
\begin{pmatrix}
c\\ -d\\ d\\\ -c
\end{pmatrix},\label{InitialState} \\ 
&\Im(c)=\Im(d)=0,\ c>0,\ d>0.\nonumber
\end{eqnarray}
These spinors correspond to $J_{x1}=-J_{x2}$, $J_{y1,y2}=J_{z1,z2}=0$ (see Appendix \ref{AppAntiferro}). In the case of $\Hcrx$, $c=1/(2\sqrt2)$, $d=\sqrt3/(2\sqrt2)$ providing the functions $\Psi^{(1,2)}_{0}$, which are the eigenfunctions of the operators $\hat J_{x1,x2}$ with the corresponding expectation values $J_{x1}=3/2$ and $J_{x2}=-3/2$. In the case of $\Hcrz$, the factors $c$ and $d$ depend on the exchange interaction and the crystal field. The crystal field interaction $\Hcrz$ leads to a partial quenching of the total magnetic moment, and the expectation values of the $\hat J_{x1,x2}$ operators are smaller than $\pm3/2$ (see Appendix \ref{AppAntiferro} and Ref.~\onlinecite{MyThesis}).

We assume, that the laser-induced Raman transitions of the atoms belonging to each sublattice involve excited states characterized by a term with $J=5/2$. Other excited states, {\it e.g.\!} with $J=3/2$ and $1/2$, are assumed to be energetically inaccessible for the applied laser pulse. It is also assumed that the Hamiltonian for the excited state is simply $\Delta_{ze}\left(3\hat J_{z1,z2}^2-\hat J_{1,2}^2\right)$ or $\Delta_{xe}\left(3\hat J_{x1,x2}^2-\hat J_{1,2}^2\right)$ and that the effect of the exchange interaction between the sublattices on the excited state is negligible. Thus, $J=5/2$ term is splitted into three doubly degenerate levels. The crystal field constants $\Delta_{ze(xe)}$ depend on the orbital radius and are not necessary equal to $\Delta_{x,z}$. We take crystal field constants for the excited states $\Delta_{ze}=3$ meV and $\Delta_{xe}=-3$ meV.  

It is assumed that all atoms belonging to the same sublattice are excited coherently by the laser pulse. Thus, the dynamics of all atoms belonging to a same sublattice can be simulated by one system. Therefore, we describe the dynamics of the antiferromagnets by considering the dynamics of the interacting sublattices 1 and 2.

\subsection{Time-dependent functions}

The functions entering the operators $\Hs^{(1)}$ and $\Hs^{(2)}$ are determined by dipole matrix elements involved in the light-induced electronic transitions between electronic states. Calculating the matrix elements, we take into account that the character of these electronic states is affected by crystal-field and magnetic interactions as described in Appendix \ref{AntiferroStatesAppendix}.

The functions $\nu_a^{(1,2)}$ and $\gamma_a^{(1,2)}$ entering the operators $\Hs^{(1)}$ and $\Hs^{(2)}$ depend on the time evolution operators $\U^{(1,2)}$, which are defined by the equation $i\,\U^{(1,2)'}=[ \hat{\mathcal H}_{\text{ex}}^{(1,2)}+\Delta_{z(x)}(3\hat J_{z1,2(x1,2)}^2-\hat J_{1,2}^2)]\,\U^{(1,2)}$, correspondingly. Since the Hamiltonians $\hat{\mathcal H}_{\text{ex}}^{(1)}$ and $\hat{\mathcal H}_{\text{ex}}^{(2)}$ are not equal, the time evolution operators $\U^{(1)}$ and $\U^{(2)}$ are not equal as well. However, the functions $\nu_a^{(1,2)}$ and $\gamma_a^{(1,2)}$ are indeed equal for both systems, $\nu_a^{(1)}=\nu_a^{(2)}=\nu_a$ and $\gamma_a^{(1)}=\gamma_a^{(2)}=\gamma_a$, due to the symmetry considerations (see Appendix \ref{AppAntiferro} and Ref.~\onlinecite{MyThesis}). Thus, the functions $\nu_a$ and $\gamma_a$ can be calculated only for the sublattice 1.

These functions are $\nu_a=\Re(Y_a)$ and $\gamma_a=\Im(Y_a)$, where $Y_a={[\U^{(1)}\mathcal A']_a}/{[\U^{(1)}\mathcal A]_a}$. The $a$-th element of $\mathcal A$ is $\mathcal A_a=1-\mathcal C_a/P_{0a}$, if $P_{0a}\neq 0$, otherwise $\mathcal A_a=0$. Here, $P_{0a}$ is the $a$-th element of $\Psi_0^{(1)}$ and $\mathcal C_a$ is the $a$-th element of the spinor
\begin{align}
&\mathcal C(t)=\mathcal E^2|d_0|^2\Biggl[\int_{-\infty}^tdt'F(t')\,\lf(\U^{(1)}\rt)^{-1}(t')\times\label{C12App}\\
&\quad\hat D^T\hat U_{\text{e}}(t')\int_{-\infty}^{t'}dt'F(t'')'\hat U_{\text{e}}(t'')\hat D\,\U^{(1)}(t'')\Biggr]
\begin{pmatrix}c\\ d\\d\\ c\end{pmatrix}\nonumber,
\end{align}
where the operator in the squared brackets acts on the initial state vector of the sublattice 1, $\Psi_0^{(1)}$. $\hat U_{\text{e}} = \exp[-i\Delta_{ze(xe)}(3\hat J_{z1(x1)}^2-\hat J_{1}^2)t]$ is the time evolution operator related to the Hamiltonian acting on the excited state and $F(t)$ is defined in Eq.~(\ref{Ft}). See Appendix \ref{SectionMomentumOpAppE} and Ref.~\onlinecite{MyThesis} for the details of the derivation and the definition of $\hat D$. 

Note that the operator $\hat{\mathcal H}_{\text{ex}}^{(1)}$, which determines $\U^{(1)}$, is time-dependent, since it depends on expectation values of $\hat J_{x2}$, $\hat J_{y2}$ and $\hat J_{z2}$ and, in general, this dependence should be taken into account. 
However, we obtain that the variation of $\hat{\mathcal H}_{\text{ex}}^{(1)}(t)$ in time, $|\hat{\mathcal H}_{\text{ex}}^{(1)}(t)-\hat{\mathcal H}_{\text{ex}}^{(1)}(0)|$, is much smaller than the laser pulse bandwidth \cite{MyThesis}. According to our calculations, the dependence of $\nu_a$ and $\gamma_a$ on the time evolution of $\hat{\mathcal H}_{\text{ex}}^{(1)}(t)$ is negligible, and the the time evolution operator can be written as $\U^{(1)}(t)=\exp[-i(\hat{\mathcal H}_{\text{ex}}^{(1)}(0)+\Delta_{z(x)}(3\hat J_{z1,2(x1,2)}^2-\hat J_{1,2}^2))t]$. The calculation of $\nu_a$ and $\gamma_a$ can be even simplified further, if the pump laser pulse bandwidth $\Delta\omega_{\text{sw}}$ is much larger than the splitting of the ground state manifold. As shown in Section \ref{SectionTimeEvolSinhleSpin} the dependence of $\nu_a$ and $\gamma_a$ on the time evolution operator $\U$ can be neglected in this case.

As discussed in Section \ref{SubsecSingleSpinEqs}, the diversity of the elements of the vector $\mathcal A$ is a necessary condition for the IFE. In the considered case, all four elements of $\mathcal A$ are different. As in the previous example, this is due to the spin-orbit coupling, which leads to dipole matrix elements of the Raman transitions from the states with different projections of the total angular momentum on the pulse propagation direction being different.

\subsection{Equations of motion}

\label{SubsecEqMotionAntiferro}


The dynamics of the magnetic vectors of the sublattices $\mathbf M_{1}=(J_{x1},J_{y1},J_{z1})$ and $\mathbf M_{2}=(J_{x2},J_{y2},J_{z2})$, can be derived from the dynamics of the expectation values of the operators $\hat N_{ab\pm}^{(1)}$ and $\hat N_{ab\pm}^{(2)}$ [see Eq.~(\ref{Nab3/2})], which act in the Hilbert space of spinorscorresponding to the sublattice 1 and 2, correspondingly. The dynamics of $\la \hat N_{ab\pm}^{(1)}\ra$ and $\la \hat N_{ab\pm}^{(2)}\ra$ can be derived from the equations of motion (\ref{EqMotionGeneral}), which must be solved for $\la N_{ab\pm}^{(1)} \ra$ and $\la N_{ab\pm}^{(2)} \ra$ simultaneously, since the systems 1 and 2 are coupled via the exchange interaction term. The action of a laser pulse on the sublattice 1 and 2 is described by operators $\Hs^{(1)}$ and $\Hs^{(2)}$, correspondingly. Note that since the mean field theory is applied and quantum fluctuations are ignored, the operators $\hat N_{ab\pm}^{(1)}$,and, consequently, the operators $\hat J_{x1}$, $\hat J_{y1}$ and $\hat J_{z1}$ commute with operators $\hat{\mathcal H}_{\text{ex}}^{(2)}$, $\Delta_z\left(3\hat J_{z2}^2-J_{2}^2\right)$, $\Hs^{(2)}$. Correspondingly, $\hat N_{ab\pm}^{(2)}$, $\hat J_{x2}$, $\hat J_{y2}$ and $\hat J_{z2}$ commute with $\hat{\mathcal H}_{\text{ex}}^{(1)}$, $\Delta_z\left(3\hat J_{z1}^2-J_{1}^2\right)$, $\Hs^{(1)}$.



We consider the equations of motion for the functions $m_{ab\pm}(t)$ and $l_{ab\pm}(t)$, which are defined as [cf.~Eq.~(\ref{ml_expval})]
\begin{align}
&m_{ab\pm}(t) = \la\hat m_{ab\pm}\ra=p_ap_b\lf\la\hat N_{ab\pm}^{(1)}+\hat N_{ab\pm}^{(2)}\rt\ra,\text{ if } b>a,\nonumber\\
&l_{ab\pm}(t) = \la\hat l_{ab\pm}\ra=p_ap_b\lf\la\hat N_{ab\pm}^{(1)}-\hat N_{ab\pm}^{(2)}\rt\ra,\text{ if } b>a, \label{defml}\\
&m_a(t)= m_{aa+}(t)=\la\hat m_a\ra=\la\hat m_{aa+}\ra=\lf\la\hat N_{a}^{(1)}+\hat N_{a}^{(2)}\rt\ra,\nonumber\\
&l_a(t)= l_{aa+}(t)=\la\hat l_a\ra=\la\hat l_{aa+}\ra=\lf\la\hat N_{a}^{(1)}-\hat N_{a}^{(2)}\rt\ra,\nonumber
\end{align}
where $a$ and $b$ are integers between 1 and 4,  $p_1=p_4=\sqrt3/2$ and $p_2=p_3=1$. This way, it is convenient to take into account symmetries of the antiferromagnetic systems. 

Additionally, instead of vectors $\mathbf M_1$ and $\mathbf M_2$, we consider the dynamics of vectors $\mathbf M=\mathbf M_1+\mathbf M_2$ and $\mathbf L = \mathbf M_1-\mathbf M_2$, which are proportional to ferromagnetic and antiferromagnetic vectors of the antiferromagnets. Instead of operators $\hat J_{\alpha}$, ($\alpha$ stays for $x$, $y$ or $z$) we use $\hat M_{\alpha} = \hat J_{\alpha 1}+\hat J_{\alpha 2}$ and $\hat L_{\alpha}=\hat J_{\alpha 1}-\hat J_{\alpha 2}$, which are connected to $\hat m_{ab\pm}$ and $\hat l_{ab\pm}$ via the relations
\begin{align}
&\hat M_x=\hat m_{12+}+\hat m_{23+}+\hat m_{34+},\nonumber\\
&\hat M_y=\hat m_{12-}+\hat m_{23-}+\hat m_{34-},\nonumber\\
&\hat M_z=\frac32\hat m_1+\frac12\hat m_2-\frac12\hat m_3-\frac32\hat m_4,\label{MLviaml}\\
&\hat L_x=\hat l_{12+}+\hat l_{23+}+\hat l_{34+},\nonumber\\
&\hat L_y=\hat l_{12-}+\hat l_{23-}+\hat l_{34-},\nonumber\\
&\hat L_z=\frac32\hat l_1+\frac12\hat l_2-\frac12\hat l_3-\frac32\hat l_4.\nonumber
\end{align}

In the Heisenberg picture, the equations of motion of the functions $m_{ab\pm}$ and $l_{ab\pm}$ are given by 
\begin{align}
&im_{ab\pm}' = \lf\l\lf[\hat m_{ab\pm},\hat{\mathcal H}_{m}+\Hs\rt]\rt\r,\nonumber\\
& il_{ab\pm}' = \l[\hat l_{ab\pm}, \hat{\mathcal H}_{m}+\Hs]\r,\label{EqmlGen}\\
&\Hs = \Hs^{(1)}+\Hs^{(2)}\nonumber.
\end{align}
There are sixteen functions $m_{ab\pm}$ and sixteen functions $l_{ab\pm}$. It is shown in Appendix \ref{AppAntiferro} that sixteen of the thirty-two functions always remain zero, namely, $m_{12\pm}(t)=0$, $m_{23\pm}(t)=0$, $m_{34\pm}(t)=0$, $m_{14\pm}(t)=0$, $l_{13\pm}(t)=0$ $l_{24\pm}(t)=0$ and $l_{1,2,3,4}(t)=0$. Applying this result together with Eq.~(\ref{EqMotionGeneral}), we obtain 
\begin{align}
m_{ab\pm}'  = &\left(-\sum_{k=1}^4\nu_km_k+\nu_a+\nu_b\right)m_{ab\pm}\nonumber\\
&\pm(\gamma_a-\gamma_b) m_{ab\mp}-i\lf\l\lf[\hat m_{ab\pm},\H\rt]\rt\r,\label{DiffEqAntiferro}\\
l_{ab\pm}'  = &\left(-\sum_{k=1}^4\nu_km_k+\nu_a+\nu_b\right)l_{ab\pm}\nonumber\\
&\pm(\gamma_a-\gamma_b) l_{ab\mp}-i\lf\l\lf[\hat l_{ab\pm},\H\rt]\rt\r\nonumber.
\end{align}
The commutators of $\hat m_{ab\pm}$ and $\hat l_{ab\pm}$ with the magnetic Hamiltonian $\H$ are given in Table \ref{CommutatorsApp} in Appendix \ref{AppAntiferro}.

Combining Eqs.~(\ref{MLviaml}) and (\ref{DiffEqAntiferro}), we obtain the following equations of motion for the vectors $\mathbf M$ and $\mathbf L$
\begin{align}
M_{x} = &0,\quad M_{y} = 0,\quad L_z = 0, \nonumber\\
L_{x}' = &F_0L_{x}+gL_{y}+F_{xy}(l_{12+},l_{34+})+G_{xy}(l_{12-}, l_{34-})\nonumber\\
&-i\lf[\hat L_x,\H\rt],\nonumber\\
L_{y}' = &F_0 L_y-gL_{x}+F_{xy}(l_{12-},l_{34-})-G_{xy}(l_{12+}, l_{34+})\nonumber\\
&-i\lf[\hat L_y,\H\rt],\label{MLdiffEqs}\\
M_{z}' =&F_0M_z+F_z -i\lf[\hat M_z,\H\rt]\nonumber,
\end{align}
where
\begin{align}
&g(t)=\gamma_2(t)-\gamma_3(t),\nonumber\\
&F_0(t)=-\sum_a^4\nu_a(t)m_a(t)+\nu_2(t)+\nu_3(t),\nonumber\\
&F_{xy}(t)(l_{12\pm},l_{34\pm})=(\nu_1(t)-\nu_3(t))l_{12\pm}(t)\nonumber\\
&\qquad+(\nu_4(t)-\nu_2(t))l_{34\pm}(t),\\
&G_{xy}(t)(l_{12\pm},l_{34\pm})=(\gamma_1(t)-2{\gamma_2}(t)+{\gamma_3}(t))l_{12\pm}(t)\nonumber\\
&\qquad+(-{\gamma_2}(t)+2{\gamma_3}(t)-{\gamma_4}(t))l_{34\pm}(t),\nonumber\\
&F_z(t)= \frac{\nu_2(t)-\nu_3(t)}{2}+(3\nu_1(t)-2\nu_2(t)-\nu_3(t))m_1(t)\nonumber\\
&\qquad+(\nu_2(t)+2\nu_3(t)-3\nu_4(t))m_4(t).\nonumber
\end{align}
Analogously to the system of differential equations in Eq.~(\ref{SpinEqMot}) describing the dynamics of the single-spin system, Eq.~(\ref{MLdiffEqs}) contains a linear term determined by the function $g(t)$ describing a rotation of the magnetic vectors around the $z$ axis. Also, analogously to the single-spin system, the terms determined by the function $F_0$ describe a rotation around the $y$ axis, which is perpendicular to the light propagation direction and to the initial alignment of the magnetic vectors. These terms are also nonlinear, since they depend on the variables $m_a$. The terms $F_{xy}$, $G_{xy}$ and $F_z$ do not appear in the equations for the single-spin system.

The set of the first-order differential equations in Eq.~(\ref{MLdiffEqs}) is not sufficient to obtain the time evolution of $M_{z}$, $L_{x}$ and $L_{z}$, since apart from these variables, the functions $m_{ab\pm}$ and $l_{ab\pm}$ also enter these differential equations. Thus, in contrast to the single-spin system, it is not possible to describe dynamics of a system with the total angular momentum $J=3/2$ induced by the UIFE with differential equations, which include only $J_x$, $J_y$ and $J_z$ to a first order (or $M_{x,y,z}$ and $L_{x,y,z}$ in the considered case). Even in the absence of the optical excitation, the dynamics of the considered antiferromagnetic systems could not be described by first-order differential equations, which include solely $M_{x,y,z}$ and $L_{x,y,z}$ variables, due to the presence of the crystal field. Thus, rather than applying Eq.~(\ref{MLdiffEqs}), it is more convenient to solve the system of differential equations in Eq.~(\ref{DiffEqAntiferro}) for variables $l_{ab\pm}$ and $m_{ab\pm}$ and to derive the time evolutions of vectors $\mathbf M$ and $\mathbf L$ with Eq.~(\ref{MLviaml}). In our case, it is sufficient to solve a system of fifteen first order differential equations, which involves fifteen variables $l_{12\pm}$, $l_{23\pm}$, $l_{34\pm}$, $l_{14\pm}$, $m_{13\pm}$, $m_{24\pm}$, $m_1$, $m_2$, $m_3$. The sixteenth variable $m_4$ is derived from the constrain $\sum_a^4m_a=\sum_a^4\la \hat N^{(1)}_a\ra+\sum_a^4\la \hat N^{(2)}_a\ra=2$.



\begin{figure}[t]
\centering
  \subfloat[]{\label{InitialVectors}\includegraphics{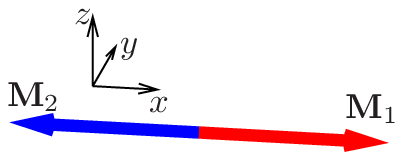}}\qquad
  \subfloat[]{\label{CircularMode}\includegraphics{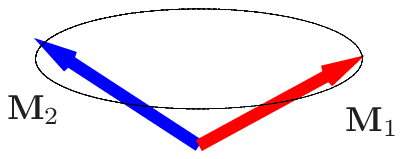}}\qquad
  \subfloat[]{\label{EllipticzMode}\includegraphics{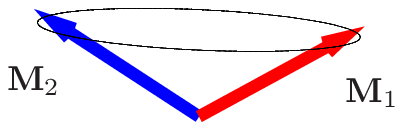}}\qquad
  \subfloat[]{\label{EllipticxMode}\includegraphics{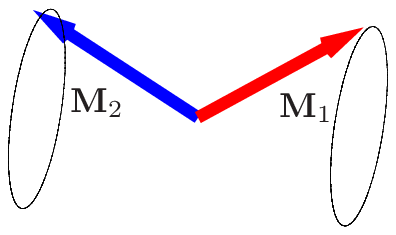}}
  \caption{(a) The initial alignment of the magnetic vectors. (b) The circular mode due to the exchange interaction. The elliptical modes due to the crystal field with the symmetry along (c) $z$ axis and (d) $x$ axis.}
\label{Modes}
\end{figure}

\subsection{Time evolution of the magnetic vectors}
\label{Subsec_timeev_magvectors}

According to Eq.~(\ref{MLdiffEqs}), $M_{x} = 0$, $M_{y} = 0$ and $L_z = 0$. Thus, the action of the UIFE makes the magnetic vectors of the sublattices $\mathbf M_1$ and $\mathbf M_2$ deviate from their equilibrium positions in such a way that their $x$ and $y$ components are opposite and $z$ components are equal [see Fig.~\ref{Modes}]. Since the magnetic vectors are deviated from their equilibrium positions, precession modes due to the exchange interaction and the crystal field are evoked in the antiferromagnetic system as shown in Fig.~\ref{Modes}(b)-(d). The exchange interaction leads to the circular rotation of the magnetic vectors around the $z$ axis  [see Fig.~\ref{Modes}(b)]. The crystal field with the symmetry in the $z$ direction leads to elliptical rotation of the magnetic vectors around the $z$ axis [see Fig.~\ref{Modes}(c)]. However, this elliptical mode exists only in the presence of the exchange interaction. The crystal field with the symmetry in the $x$ direction leads to elliptical rotation of the magnetic vectors around the $x$ axis [see Fig.~\ref{Modes}(d)] in such a way that their $x$ and $y$ projections are always opposite to each other, and $z$ projections are alway equal to each other.



\begin{figure}[]
\centering       
\includegraphics[width=0.44\textwidth]{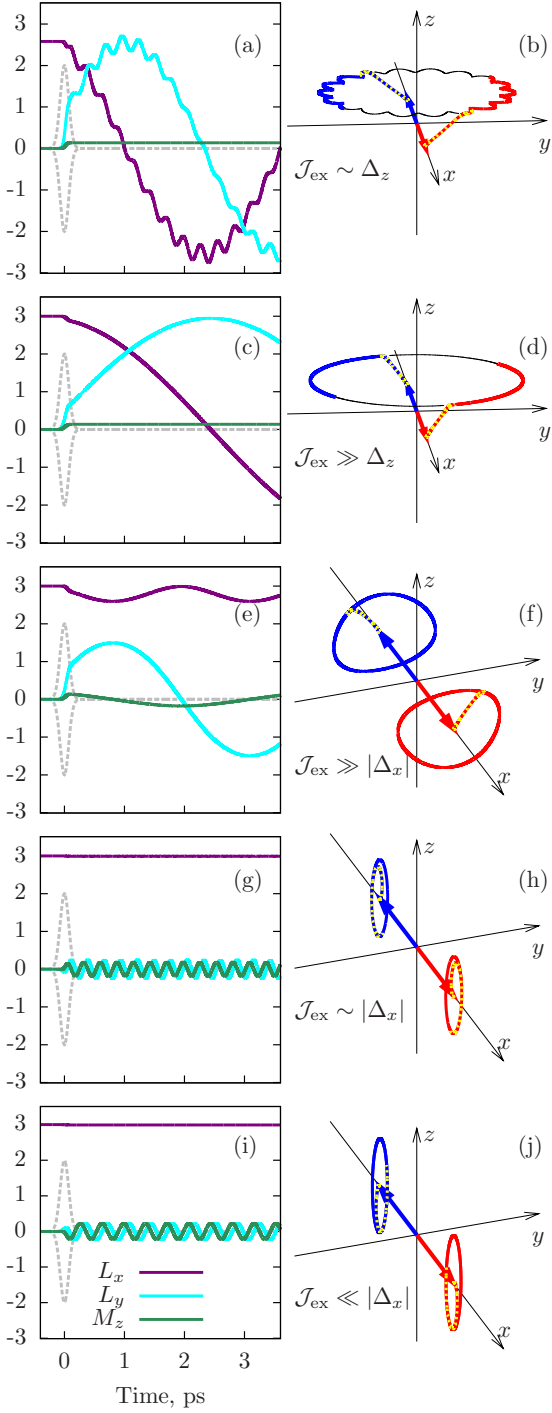}
\caption{Left column: the time evolutions of $L_x$, $L_y$ and $M_z$ depending on the values of $\J$ and  $\Delta_{x,z}$. The gray line represents the time evolution of the electric field amplitude. Right column: the corresponding trajectory of the $\mathbf M_1$ (red) and $\mathbf M_2$ (blue) vectors. The dotted yellow line represents the dynamics of the vectors during the excitation. (a),(b) $\J=3$ meV, $\Delta_z = 2$ meV. (c),(d) $\J=3$ meV, $\Delta_z = 0.02$ meV. (e), (f) $\J=3$ meV, $\Delta_x = 0.02$ meV.  (g), (h) $\J=3$ meV, $\Delta_x = -2$ meV. (i), (j) $\J\approx0$ meV,  $\Delta_x = -2$ meV.}
%
\label{Dynamics}
\end{figure}

We calculated the time evolutions of the magnetic vectors triggered by a left-circularly polarized Gaussian shaped laser pulse of the duration of 117 fs, which corresponds to bandwidth of 15 meV. We assume the peak intensity of $2\times10^{10}$ W/cm$^2$,  the fluence 8 mJ/cm$^2$ and the central frequency of 2.0 eV. Figure~\ref{Dynamics} shows the time evolution of the components $L_x = M_{1x}-M_{2x}$, $L_y = M_{1y}-M_{2y}$ and $M_z = M_{1z}+M_{2z}$ and the corresponding 3D picture of the trajectories of the magnetic vectors $\mathbf M_1$ and $\mathbf M_2$ at different values of the exchange interaction and the crystal field interaction constants. The red and blue arrows on the right panel show the initial alignments of the magnetic vectors of the sublattices 1 and 2 correspondingly. The dotted yellow-red and yellow-blue lines show the dynamics of the corresponding vectors during the excitation. The continuous red and blue lines show their dynamics after the excitation. 

The dynamics of the magnetic vectors during and after the action of the laser pulse depends considerably on the exchange and crystal field interactions. Figures \ref{Dynamics}(a)-(d) show the dynamics of the magnetic vectors in the case of the crystal field $\Hcrz$ with the symmetry along the $z$ axis. In this case, vectors $\mathbf M_1$ and $\mathbf M_2$ move upwards during the action of the pump pulse and start to precess around the $z$ axis slightly before the excitation finishes. Their $z$ projections remain constant after the excitation, since the oscillation modes due to the exchange interaction and the crystal field $\Hcrz$ correspond to constant $M_z$ [cf.~Figs.~\ref{Modes}(b) and (c)]. The precession of the magnetic vectors involves both elliptical and circular modes, when $\mathcal J_{\text{ex}}\sim \Delta_z$ [cf.~Figs.~\ref{Dynamics}(a) and (b)]. The length of the magnetic vectors is slightly lower than 3/2 due to the partial quenching of the angular momentum by the crystal field as discussed earlier. The magnetic vectors simply circulate around the $z$ axis, when $\mathcal J_{\text{ex}}\gg \Delta_z$ [cf.~Figs.~\ref{Dynamics}(c) and (d)].

In the case of the crystal field $\Hcrx$ with the symmetry along the $x$ axis, the dynamics of the magnetic vectors during the excitation are much more dependent on the value of the crystal field constant [cf.~Figs.~\ref{Dynamics}(e)-(j)]. After the excitation, the magnetic vectors start to follow an elliptical trajectory around the $x$ axis, which is slightly bent to the $z$ axis in the case of $\J\gg|\Delta_x|$ [cf.~Fig.~\ref{Dynamics}(e)-(f)]. As discussed earlier, the magnetic vectors always move in such a way that their projections on the $z$ axis are equal to each other, and the $x$ and $y$ projections are opposite.

The trajectories of the magnetic vectors during the excitation in the case of $\J\gg |\Delta_x|$ [cf.~Figs.~\ref{Dynamics}(e) and (f)] are quite similar to the the ones in the case of $\J\sim\Delta_z$ [cf.~Figs.~\ref{Dynamics}(a) and (b)] and $\J\gg\Delta_z$ [cf.~Figs.~\ref{Dynamics}(c) and (d)]. Magnetic vectors on Figs.~\ref{Dynamics}(e) and (f) also start to precess slightly before the excitation has finished. This means that the time evolutions of the magnetic vectors during the excitation are not strongly influenced by the exchange and crystal field interactions, and approximate positions of the magnetic vectors after the excitation are determined mainly by the interaction with the pump pulse. Still, it is noticeable that the values of $L_y$ right after the excitation on Figs.~\ref{Dynamics}(a),(c) and (e) differ from each other. In all these cases, the periods of the induced precessions are more than twenty times larger than the pulse duration.

A situation is quite different in the case of $\J\sim |\Delta_x|$ and $\J\ll |\Delta_x|$. The magnetic vectors start to precess during the excitation at approximately half of the pulse duration [cf.~Figs.~\ref{Dynamics}(g)-(j)]. The trajectories during the excitation are strongly dependent on the value of the crystal field constant $\Delta_x$. The magnetic vectors even move downwards before the start of the precession on Figs.~\ref{Dynamics}(i) and (j)], which is the opposite direction to the ones in all other cases in Fig.~\ref{Dynamics}. This means that the dynamics of the magnetic vectors during the action of the pump pulse is strongly dependent on the exchange interaction and the crystal field $\Hcrx$. The period of the induced precessions is about two times longer than the pulse duration, when $\J\sim |\Delta_x|$, and about three time longer than the pulse duration, when $\J\ll |\Delta_x|$.

Thus, if the pulse duration is several tens of times shorter than the period of laser-induced magnetic precessions, then the trajectories of the magnetic vectors during the excitation are similar even at different values of the crystal-field constants of the antiferromagnetic system. However, the positions of the magnetic vectors right after the excitation still slightly differ in this case. If the pulse duration is just several times shorter than the period of the laser-induced magnetic precessions, then the dynamics of the magnetic vectors during the excitation can be absolutely divergent at different values of the exchange and crystal field constants. Thus, the dynamics of the magnetic vectors during the excitation are mainly determined by the action of the pump pulse in the former case. In the latter case, the magnetization dynamics during the excitation are mainly determined by the internal magnetic interactions and the pump pulse serves as a slight impulse prompting the dynamics. This demonstrates that an accurate calculation of the magnetization dynamics during the action of light is necessary to predict correct positions of the magnetic vectors right after the excitation.


\section{Conclusions}

The action of the UIFE on a magnetic system leads to deviation of its magnetic moment from the ground state, which prompts the magnetic moment to precess \cite{MyThesis}. In our article, we have shown that this process cannot be described within the sudden approximation even if the pump-pulse duration is several tens of times shorter than the period of laser-induced precessions. We provide a technique that accurately describes magnetization dynamics during the action of a laser pulse at a subpicosecond time scale.

We derived the Heisenberg representation for the UIFE from the Schr\"odinger picture, which describes coupling of light to electrons of a magnetic system. We obtained an operator $\Hs$ acting in the Hilbert space of total angular momentum with time-dependent elements, which depend on laser-pulse parameters and transition amplitudes of the electronic system under the action of a laser pulse. This way we substituted the operator $-\mathbf d\cdot\mathbf E$ by the nondiagonal effective magnetic operator $\Hs$. The effective magnetic operator allows to separate the motion of a magnetic vector due to the action of light from that induced by other fields acting on a magnetic system. Commuting the magnetic operator with total angular momentum operators, we obtained equations of motion for magnetic vectors of a magnetic system. During the action of light, magnetic vectors move due to the joint action of $\Hs$ and other magnetic operators acting on the magnetic system. After the action of light, the elements of $\Hs$ naturally become zero.

The Heisenberg representation of the UIFE could be implemented for macroscopic calculations of laser-induced magnetization dynamics, which are a practical technique allowing to take simultaneously many different magnetic effects into account \cite{KirilyukRMP10, AtxitiaPRB11, NievesProceedings15, MorenoPRB17}. The effective magnetic operator $\Hs$ could be also used as a convenient tool to adjust pump-pulse properties enhancing the UIFE, since it directly illustrates how a pump pulse couples to the total angular momentum.

With the help of the illustrative single-spin system in an external magnetic field, we showed that laser-induced magnetization  dynamics can be strongly affected by the Larmor precession even during the action of a laser pulse. The spin started to precess during the action of the laser pulse even when the Larmor period was forty times longer than the pump-pulse duration. Thus, even in this case, it was necessary to calculate the joint action of the external magnetic field and the UIFE on spin in order to obtain the correct position of the spin vector after the excitation.

We calculated magnetization dynamics induced by the UIFE in model antiferromagnetic systems consisting of two sublattices with opposite magnetic vectors. The magnetization dynamics in these systems were described by a system of fifteen first-order differential equations within the Heisenberg picture. We demonstrated that the action of the UIFE induced by a pump pulse propagating in the $z$ direction made both magnetic vectors bend upwards to the $z$ axis and rotate around it in such a way that their $z$ projections were equal, and $x$ and $y$ projections were opposite. The deviation of the magnetic vectors from their initial positions evoked a circular precession mode due to the exchange interaction and an elliptical precession mode due to the crystal field. 

In the case of the $xy$-easy-plane antiferromagnetic system with the exchange-interaction constant of 3 meV and the crystal-field constant of 2 meV, the period of the induced precessions was more than twenty times longer than the pump-pulse duration. The motion of the magnetic vectors was mainly determined by the UIFE and to some extend affected by the crystal field and exchange interactions. In the case of the $z$-easy-axis antiferromagnetic system with the same absolute values of the exchange interaction and crystal-field constants, the period of the induced precessions was just two times longer than the pump-pulse duration. The motion of the magnetic vectors was mainly determined by the crystal field and the exchange interaction, and the pump pulse served just as a slight impulse prompting the magnetization dynamics. The magnetic vectors moved even in the opposite direction to the one, in which the action of the UIFE alone would make the magnetic vectors move. This example also demonstrates that an accurate calculation of magnetization dynamics during the action of a pump pulse is necessary even if the pump-pulse duration is several tens of times shorter than the period of induced magnetic precessions. 

We thus observed that the character of induced magnetic precessions depends on the ratio of the pump-pulse duration to the period of magnetic-oscillation modes. This effect could be used for control of magnetic precessions by varying the pump-pulse duration in the regime, when it is comparable with the period of magnetic oscillation modes. In this regime, the motion due to the UIFE-driven deviation of magnetic vectors can cooperate or compete with the one due to magnetic precessions caused by this deviation. For example, in the $z$-easy-axis antiferromagnetic system, the UIFE-induced deviation is in the upward $z$ direction, but the crystal-field interaction causes the precession upwards and downwards the $z$ axis. Varying the pump-pulse duration, one can apply the driving during the time, when it either enhances or counteracts the precession mode, which would influence the magnitude and the phase of resulting precessions after the action of the pump pulse. This effect does not follow from the phenomenological description of the IFE, in which the pump-pulse fluence alone determines the character of magnetic precessions.

The developed technique to study the magnetization dynamics induced by the UIFE can be applied to other types of magnetic materials, not necessary antiferromagnetic. The Heisenberg representation of other magneto-optical effects driven by Raman transitions \cite{KalashnikovaPRL07, NishitaniAPL10, KandaNature11, TzschaschelPRB17, KhanPRB20} can be derived analogously to our methodology. The presented concept of the time-dependent effective magnetic operator paves the way towards the macroscopic description of ultrafast laser-induced magnetization dynamics that accurately takes electronic transitions induced by an ultrashort light pulse into account.

\section*{Acknowledgment}
This work was supported by European Community's Seventh Framework Programme FP7/2007-2013 under Grant Agreement No.~214810 (FANTOMAS). D.~P.-G. acknowledges the support of the Volkswagen Foundation.

\begin{appendix}

\section{Effective magnetic operator describing the UIFE}
\label{MomentumOperatorAppendix}
\subsection{Derivation of the effective magnetic operator}

Let us first derive the operator for the case, when there is no field except light acting on the total angular momentum of the system, meaning that $\U=\mathbf 1$ and $i\Psi'_g=\Hs^0\Psi^0_g$. Here, $\Hs^0$ and $\Psi^0_g$ are the operator and the time-dependent spinor in this case. The relation (\ref{Adef}) can be rewritten as
\begin{align}
&\Psi^0_g=\frac{\mathcal A\Psi_0}{\norma}=
\frac{1}{\norma(t)}
\begin{pmatrix}
A_1(t)e^{i\phi_1(t)}P_{01}\\\vdots\\A_k(t)e^{i\phi_k(t)}P_{0k}\\\vdots
\end{pmatrix}=
\begin{pmatrix}
P_{1}(t)\\\vdots\\P_{k}(t)\\\vdots
\end{pmatrix},
\end{align}
where the elements of the spinor $\mathcal A$ are $\mathcal A_k = A_k(t)e^{i\phi_k(t)}$.

We apply the condition that the magnetic moment of a system is not rotated by the IFE, if it is parallel to the laser pulse propagation direction. The magnetic moment is parallel to the quantization axis, if all elements except one of the spinor $\Psi_0$ are zero. This means that, if $\Psi_0=\left(\begin{smallmatrix}0\\ \vdots\\ P_{0k}\\ \vdots\end{smallmatrix}\right)$, there $|P_{0k}|=1$, then the effective magnetic operator acts only on the $k$-th component of the state vector, so that the other magnetic components remain zero, $\Psi^0_g=\frac1{\norma(t)}\left(\begin{smallmatrix}0\\ \vdots\\ A_k(t)e^{i\phi_k(t)}\\ \vdots\end{smallmatrix}\right)$. Since $|\Psi^0_g|=1$, the spinor during and after the action of light is $\Psi^0_g=e^{i\phi_k(t)}\Psi_0$. Thus, the diagonal elements of $\Hs^0$ are $(\Hs^0)_{aa}=-\phi'_a(t)$, and if the magnetic moment is parallel to the light propagation direction, the off-diagonal elements of $\Hs^0$ must become zero.

Let us now derive the off-diagonal elements of $\Hs^0$, which are non-zero, if the magnetic moment is not parallel to the light propagation direction. The $a$-th element of $i{\Psi^0_g}'$ can be expressed as
\begin{align}
iP'_a
=-\phi'_aP_a+i\sum_{b,b\neq a}P_b\left[P_aP_b^*\left(\frac {A'_a}{A_a}-\frac{A'_b}{A_b}\right)\right]\label{Psi_prime_App},
\end{align}
which follows from the relations $P_{a}(t) = A_a(t)e^{i\phi_a(t)}P_{0a}/\sum_k|P_k(t)|$ and $P_{0a}=\text{const}$. Applying this expression with $i{\Psi^0_g}'=\Hs^0\Psi^0_g$, we obtain
\begin{equation}
\Hs^0=\begin{pmatrix}
\ddots&&&&\\
&-\gamma_a&\cdots&i P_a P_b^*\left(\nu_a-\nu_b\right)&\cdots\\
&\vdots&\ddots&&\\
&i P^*_a P_b\left(\nu_b-\nu_a\right)&&&\\
&\vdots&&&
\end{pmatrix},\label{MomOpMatrixApp}
\end{equation}
where $\nu_{a}=\Re(\mathcal A_a'/\mathcal A_a)$, $\gamma_a=\phi_a'=\Im(\mathcal A_a'/\mathcal A_a)$ for $\U=\mathbf 1$.

If there is a field $\H$, which acts on the magnetic system apart from light, then $\U\neq\mathbf 1$ and $i\Psi_g'=\lf[\Hs+\H\rt]\Psi_g$. In this case the spinor is $\Psi_g={\U \mathcal A\Psi_0}/{\norma} = \U\Psi^0_g$. Substituting $\U\Psi^0_g$ for $\Psi_g$, we obtain that $i(\U\Psi^0_g)'=\lf[\Hs+\H\rt]\U\Psi^0_g$. Applying that, by definition, $i\,\U' = \H\,\U$, we obtain $\Hs = \U\,\Hs^0\,\U^{-1}$. We derive $\Hs$ applying this expression, which results in the operator $\Hs$ having the same form as $\Hs^0$ in Eq.~(\ref{MomOpMatrixApp}), but with different functions $\nu_a$ and $\gamma_a$:
\begin{equation}
\nu_a=\Re\left(Y_a\right), \quad\gamma_a=\Im\left(Y_a\right),\quad Y_a={[\U\mathcal A']_a}/{[\U \mathcal A]_a},
\end{equation}
where $[\U\mathcal A']_a$ is the $a$-th element of the spinor obtained by the action of the operator $\U$ on $\mathcal A'$, which is the time-derivative of the spinor $\mathcal A$.  $[\U \mathcal A]_a$ is the $a$-th element of the spinor obtained by the action of the operator $\U$ on the spinor $\mathcal A$. Applying that $P_aP_b^*=(\l\hat N_{ab+}\r+i\l\hat N_{ab-}\r)/2 = \l\Psi_g|\hat N_{ab+}+i\hat N_{ab-}|\Psi_g\r/2$, the effective magnetic operator can be written as
\begin{align}
\Hs=&-\sum_{a}^n\gamma_a\hat N_a\label{OperatorGeneralApp}\\
&+\frac12\sum_{a,b}^n(\nu_a-\nu_b)\bigl( \l \hat N_{ab-}\r\hat  N_{ab+}- \l \hat N_{ab+}\r \hat N_{ab-}\bigr)\nonumber.
\end{align}

\subsection{Commutator with the effective magnetic operator}

In this subsection, we derive the expectation value of the commutator $-i\lf\l\lf[\hat N_{ab\pm},\Hs\rt]\rt\r$. We divide the operator $\Hs$ into its diagonal, $H_{d}$, and off-diagonal part, $H_n$, and derive the expectation values of the commutators $-i\lf\l\lf[\hat N_{ab\pm},H_{d}\rt]\rt\r$ and $-i\lf\l\lf[\hat N_{ab\pm},H_{n}\rt]\rt\r$ separately. 

The $cd$-th matrix element of the commutator of $\lf[\hat N_{ab\pm},H_{d}\rt]$ is
\begin{align}
&\lf(\hat N_{ab\pm}\hat H_{d}-\hat H_{d}\hat N_{ab\pm}\rt)_{cd}=(\hat N_{ab\pm})_{cd}\lf[(\hat H_{d})_{dd}-(\hat H_{d})_{cc}\rt],
\end{align}
where we designate the $cd$-th matrix element of an operator $\hat O$ as $(\hat O)_{cd}$. Thus, $\lf[\hat N_{ab\pm},\hat H_d\rt]=\pm i(\gamma_a-\gamma_b)\hat N_{ab\mp}$. 

Let us now commute $\hat N_{ab\pm}$ with the off-diagonal part $\hat H_{n}$ and determine the expectation value of the commutator $-i\l[\hat N_{ab\pm},\hat H_{n}]\r$. We consider the elements of the commutator $\hat C^{ab\pm}=[\hat N_{ab\pm},\hat H_{n}]$. We find that $(\hat C^{ab\pm})_{cd}=0$ if neither $c$ nor $d$ are not equal $a$ or $b$. The elements in other cases are
\begin{align}
&(\hat C^{ab\pm})_{ac}=(\hat N_{ab\pm})_{ab}(\hat H_{n})_{bc},\nonumber\\
& (\hat C^{ab\pm})_{ca}=-(\hat N_{ab\pm})_{ba}(\hat H_{n})_{cb},\label{App_ElemComm}\\
&(\hat C^{ab\pm})_{bc}=(\hat N_{ab\pm})_{ba}(\hat H_{n})_{ac},\nonumber\\
& (\hat C^{ab\pm})_{cb}=-(\hat N_{ab\pm})_{ab}(\hat H_{n})_{ca}\nonumber
\end{align}
The expectation value $\l \hat C^{ab\pm}\r = \l \Psi_g|\hat C^{ab\pm}|\Psi_g\r$ is given by $\sum_{cd}P^*_cP_d\hat C^{ab\pm}_{cd}$, which is equal to $i[-2(\sum_i|P_i|^2\nu_i)+(\nu_a+\nu_b)](P_aP_b^*\pm i P_a^*P_b)$ as follows from Eq.~(\ref{App_ElemComm}). Applying that $|P_a|^2=\l\Psi_g|\hat N_a|\Psi_g\r$ and $(P_aP_b^*\pm i P_a^*P_b)=\l \hat N_{ab\pm}\r$, we obtain the relation
\begin{align}
-i\lf\l\lf[\hat N_{ab\pm},\Hs\rt]\rt\r = &\lf(-2\sum_k\nu_k \l \hat N_k\r+\nu_a+\nu_b\rt) \l \hat N_{ab\pm} \r\nonumber\\
&\pm\lf(\gamma_a-\gamma_b\rt) \l \hat N_{ab\mp}\r.
\end{align}

\section{Spin-orbit coupling and Zeeman interaction}

\label{AppMagField}

In order to obtain the wave functions and the splitting of the $2p$ state in the presence of the spin-orbit coupling and Zeeman interaction, one has to diagonilize the following integral is
\begin{equation}
\H=-\frac{B}{2}(2\hat S_x+\hat L_x)-\SOC\mathbf L\cdot\mathbf S.
\end{equation}
This Hamitonian has six eigenvectors and eigenenergies for the $2p$ state. Thus, $2p$ state is split energetically into six states with energies $\varepsilon_{k_\pm}=\varepsilon_{2p,k_\pm}+E_{k_\pm}$, $k$ = 1, 2 or 3. The indices $k_\pm$ correspond to the indices $j$ in Section \ref{SubsecSingleSpinEqs}. $E_{k_\pm}$ are the solutions of the equations
\begin{align}
&E_{k_\pm}^3\pm\frac{B}{2}E_{k_\pm}^2-\left(\frac{3\SOC^2}{4}\pm\frac{B\SOC}{2}+\frac{B^2}{2}\right)E_{k_\pm}\\
&+\left(-\frac{\SOC^3}{4}+\frac{\SOC B^2}{4}\mp\frac{3\SOC^2 B}{8}\right)=0\nonumber.
\end{align}
The corresponding wave functions are, if $\SOC\neq0$ and $B\neq0$,
\begin{align}
&\Psi^{2p}_{k_\pm}=\alpha_{k_\pm}\bigl(|L_z=1,S_z=1/2\r \pm |L_z=-1,S_z=-1/2\r\bigr)\nonumber  \\
&+\beta_{k_\pm}\bigl(|L_z=1,S_z=-1/2\r \pm |L_z=-1,S_z=1/2\r\bigr)\label{WavefMagSOCApp}\\
&+\gamma_{k_\pm}\bigl(|L_z=0,S_z=1/2\r\pm |L_z=1,S_z=-1/2\r\bigr),\nonumber
\end{align}
where $\alpha_{k_\pm}={B\left(5\SOC/8-3E_{k_\pm}/4\right)}/[{\norma_{k_\pm}\left(E_{k_\pm}+\SOC/2\right)}]$, $\beta_{k_\pm}=(E_{k_\pm}+B/4-\SOC/2)/{\norma_{k_\pm}}$ and $\gamma_{k_\pm} = (E_{k_\pm}-3\SOC/2- B/2)/({\sqrt2\norma_{k_\pm}})$ and
$\norma_{k_\pm}$ is the normalisation factor, which provides $|\alpha_{k_\pm}|^2+|\beta_{k_\pm}|^2+|\gamma_{k_\pm}|^2=1$.

The dipole matrix elements for a transition from the $s$ state with the wave function $|L_z=0,S_z=1/2\r$ to the states $k_\pm$, $d_{\uparrow k_\pm}$, are proportional to $\alpha_{k_\pm}$ for the left-circularly polarized light. The dipole matrix elements for a transition from the $s$ state with the wave function $|L_z=0,S_z=-1/2\r$ to the states $k_\pm$, $d_{\downarrow k_\pm}$, are proportional to $\beta_{k_\pm}$ for the left-circularly polarized light.

If $\SOC=0$ and $B\neq0$, then $\alpha_{1_\pm} = \beta_{1_\pm} = 1/(2\sqrt2)$, $\alpha_{2_\pm} = -\beta_{2_\pm} = 1/2$, $\alpha_{3_\pm} =\beta_{3_\pm}= 1/(2\sqrt2)$. Thus, if $\SOC = 0$, but $B\neq0$, no rotation is possible, since $|d_{\uparrow k_\pm}|^2 = |d_{\downarrow k_\pm}|^2$ [see Eq.~(\ref{AforSpin1/2})].

\section{Antiferromagnetic system}
\label{AppAntiferro}

\subsection{Ground and excited states}
\label{AntiferroStatesAppendix}

In Section \ref{AntiferroStatesCh6}, we consider the following light field-free magnetic Hamiltonians acting on the sublattices 1 and 2, correspondingly,
\begin{align}
\H^{(1)}=&\J(J_{x2}\hat J_{x1}+J_{y2}\hat J_{y1}+J_{z2}\hat J_{z1})\nonumber\\
&+\Delta_{z(x)}\left(3\hat J_{z1(x1)}^2-\hat J_1^2\right)\\
\H^{(2)}=&\J(J_{x1}\hat J_{x2}+J_{y1}\hat J_{y2}+J_{z1}\hat J_{z2})\nonumber\\
&+\Delta_{z(x)}\left(3\hat J_{z2(x2)}^2-\hat J_2^2\right)
\end{align}
The sign of the crystal field constants $\Delta_{z}$ and $\Delta_{x}$ are chosen in such a way that the alignment of magnetic vectors of the sublattices along the $x$ axis is energetically preferable. Due to the exchange interaction, the state with the lowest energy corresponds to the magnetic vectors having the largest possible amplitude and being opposite to each other: $J_{x1}=-J_{x2}$.  Thus, the effective magnetic Hamiltonians before the action of light can be represented as
\begin{align}
&\hat{\mathcal H}_{m\text{(eff)}}^{(1)} = \Delta_{z(x)}\left(3\hat J_{z1(x1)}^2-\hat J_1^2\right)+\Jx\hat J_{x1},\label{H0eff}\\
&\hat{\mathcal H}_{m\text{(eff)}}^{(2)} = \Delta_{z(x)}\left(3\hat J_{z2(x2)}^2-\hat J_2^2\right)+\Jx\hat J_{x2}\nonumber
\end{align}
$\Jx= \J J_{x1} = - \J J_{x2}$. 

The ground state spinors, corresponding to the following expectation values of total angular momentum operator $J_{x1}=-J_{x2}$,  $J_{x1}>0$, $J_{y1}=J_{y2}=J_{z1}=J_{z2}=0$, must fulfill the conditions $\lf\l\Psi^{(1)}_{0}\lf|\hat J_{x1}\rt|\Psi^{(1)}_{0}\rt\r=-\lf\l\Psi^{(2)}_{0}\lf|\hat J_{x2}\rt|\Psi^{(2)}_{0}\rt\r$ and $\lf\l\Psi^{(1,2)}_{0}\lf|\hat J_{y1,y2}\rt|\Psi^{(1,2)}_{0}\rt\r=\lf\l\Psi^{(1,2)}_{0}\lf|\hat J_{z1,z2}\rt|\Psi^{(1,2)}_{0}\rt\r=0$, where
\begin{align}
&\hat J_{x1,x2}=
\begin{pmatrix}
0&\frac{\sqrt3}{2}&0&0\\
\frac{\sqrt3}{2}&0&1&0\\
0&1&0&\frac{\sqrt3}{2}\\
0&0&\frac{\sqrt3}{2}&0
\end{pmatrix}, \\
&\hat J_{y1,y2}=
\begin{pmatrix}
0&-\frac{i\sqrt3}{2}&0&0\\
\frac{i\sqrt3}{2}&0&-i&0\\
0&i&0&-\frac{i\sqrt3}{2}\\
0&0&\frac{i\sqrt3}{2}&0
\end{pmatrix},\\
&\hat J_{z1,z2}=
\begin{pmatrix}
\frac32&0&0&0\\
0&\frac12&0&0\\
0&0&-\frac12&0\\
0&0&0&-\frac32
\end{pmatrix}.
\end{align} 
The spinors, which fulfill these conditions, have the following form
\begin{eqnarray}
&\Psi^{(1)}_{0}=
\begin{pmatrix}
c\\d\\ d\\ c
\end{pmatrix},\ 
\Psi^{(2)}_{0}=
\begin{pmatrix}
c\\ -d\\ d\\\ -c
\end{pmatrix},\label{InitialState} \\ 
&\Im(c)=\Im(d)=0,\ c>0,\ d>0.\nonumber
\end{eqnarray} 

In the case of the crystal field $\Hcrx$, the effective Hamiltonians are diagonal in the basis with the quantization axis along the $x$ axis. Thus, the ground state state vectors are the eigenvectors of the $\hat J_{x1,x2}$ operators. The ground state spinors correspond to the expectation values $J_{x1}=3/2$ and $J_{x2}=-3/2$ with $c=1/(2\sqrt2)$, $d=\sqrt3/(2\sqrt2)$.

The situation is more complicated in the case of $\Hcrz$. The effective Hamiltonian is not diagonal in a basis with neither $x$ nor $z$ quantization axes. The spinors depend on the crystal field and the exchange interaction, and must be found numerically. If $\Delta_z\neq 0$, then the expectation values have lower values $J_{x1}<3/2$ and $J_{x2}>-3/2$, and the crystal field $\Hcrz$ leads to a partial quenching of the magnetic moment.

\subsection{Time-dependent factors of the effective magnetic operator}

\label{SectionMomentumOpAppE}
In this subsection we calculate the functions $\nu_a^{(1,2)}$ and $\gamma_a^{(1,2)}$ and show that they are equal for both systems. These functions are defined as $\nu_a^{(1,2)}=\Re(Y_a^{(1,2)})$, $\gamma_a^{(1,2)}=\Im(Y_a^{(1,2)})$, where $Y_a={[\U^{(1,2)}{\mathcal A^{(1,2)}}']_a}/{[\U^{(1,2)}\mathcal A^{(1,2)}]_a}$. The spinors $\mathcal A^{(1,2)}$ are defined by Eq.~(\ref{Adef}). $\U$ is the time evolution operator defined by $i\,\U^{(1,2)'}=\hat{\mathcal H}_m^{(1,2)}\, \U^{(1,2)}$. 

According to Eqs.~(\ref{ExpansionInt}), (\ref{Psig}) and (\ref{Adef}) the $a$-th elements of $\mathcal A_{1,2}$ is $\mathcal A_a^{(1,2)}=1-\mathcal C_a^{(1,2)}/P_{0a}^{(1,2)}$, if $P_{0a}^{(1,2)}\neq 0$, otherwise, $\mathcal A_a^{(1,2)}=0$. 
Here, $P_{0a}^{(1,2)}=P_{a}(0)^{(1,2)}$, $\mathcal C_a^{(1,2)}$ is the $a$-th element of the vector obtained by the action of the operator in squared brackets on the initial spinor
\begin{align}
\mathcal C^{(1,2)}=&\Biggl[\int_{-\infty}^tdt'\,(\U^{(1,2)})^{-1}\hat V\U^{(1,2)}\nonumber\\
&\times\int_{-\infty}^{t'}dt''(\U^{(1,2)})^{-1}\hat V\U\Biggr]\label{VectorCApp}
\Psi_0^{(1,2)}.
\end{align}

Since the action of the exchange interaction on the excited state is ignored, the time evolution operator, which acts on spinor of the excited state, can be written simply as an operator $\hat U_{\text{e}}(t)=e^{-i\hat{\mathcal H}_{\text{cr}z(\text{cr}x)}t}$. It is equal for the both systems 1 and 2. The crystal field operator, $\Hcrz$, acting on the excited state characterized by the term $J=5/2$, is represented by 
\begin{align}
\Hcrz&=\Delta_{ze}\left(3\hat J_{z1}^2-\frac{35}4\right)\\
&=
\Delta_{ze}\begin{pmatrix}
10& 0& 0& 0& 0& 0\\
0& -2 & 0 &0& 0& 0\\
0& 0 & -8& 0 &0\\
0& 0 & 0 & -8& 0&0\\
0& 0& 0& 0&-2 & 0\\
0& 0& 0 &0& 0& 10
\end{pmatrix}\nonumber
\end{align}
Thus, the crystal field leads for both systems to the splitting of the excited state $J=5/2$ into the following three doubly degenerate states with the corresponding energies $\varepsilon_{\text{ex1,2,3}}$
\begin{align}
&|J_{z1(z2)}=\pm5/2\rangle,\ \varepsilon_{\text{ex1}}=\varepsilon_{\text{ex}}+10\Delta_{ze}\nonumber\\
&|J_{z1(z2)}=\pm3/2\rangle,\ \varepsilon_{\text{ex2}}=\varepsilon_{\text{ex}}-2\Delta_{ze}\\
&|J_{z1(z2)}=\pm1/2\rangle,\ \varepsilon_{\text{ex3}}=\varepsilon_{\text{ex}}-8\Delta_{ze}\nonumber,
\end{align} 
where $\varepsilon_{\text{ex}}=2$ eV is the energy of the excited state in the absence of the crystal field. 

The crystal field operator, $\Hcrx$, acting on the excited state characterized by the term $J=5/2$, is represented by
\begin{align}
\Hcrx&=\Delta_{xe}\left(3\hat J_{x1}^2-\frac{35}4\right)\\
&=
\Delta_{xe}\begin{pmatrix}
-5& 0& \frac{3\sqrt10}{2}& 0& 0& 0\\
0& 1 & 0 &\frac{9\sqrt2}{2}& 0& 0\\
\frac{3\sqrt10}{2}& 0 & 4& 0 & \frac{9\sqrt2}{2}\\
0& \frac{9\sqrt2}{2} & 0 & 4& 0&\frac{3\sqrt10}{2}\\
0& 0& \frac{9\sqrt2}{2}& 0&1 & 0\\
0& 0& 0 &\frac{3\sqrt10}{2}& 0& -5
\end{pmatrix}.\nonumber
\end{align}
The crystal field leads for both systems to the splitting of the excited state $J=5/2$ into the following three doubly degenerate states with the corresponding energies $\varepsilon_{\text{ex1,2,3}}$
\begin{align}
&|J_{x1(x2)}=\pm1/2\rangle,\ \varepsilon_{\text{ex1}}=\varepsilon_{\text{ex}}-8\Delta_{xe},\nonumber\\
&|J_{x1(x2)}=\pm3/2\rangle,\ \varepsilon_{\text{ex2}}=\varepsilon_{\text{ex}}-2\Delta_{xe},\\
&|J_{x1(x2)}=\pm5/2\rangle,\ \varepsilon_{\text{ex3}}=\varepsilon_{\text{ex}}+10\Delta_{xe}.\nonumber
\end{align}


Let us examine the selection rules for the transitions from the ground state $|\text{g}\r$ to the excited state $|\text{ex}\r$ for an excitation by left-circularly polarized light. A dipole matrix element of a transition from a state with a total angular momentum $J$ and projection $J_z=m$ to a state with $J+1$ and $J_z=m+1$ is\cite{LandauQuantum}
\begin{align}
&\langle J+1\,m+1|r_+|J\,m\rangle=\\
&-\sqrt\frac{(J+m+1)(J+m+2)}{(J+1)(2J+1)(2J+3)}\,\langle J+1|r|J\rangle\nonumber.,
\end{align}
where $r_+=(x+iy)/\sqrt{2}$. Thus, the action of the operator $\hat V$ on the ground state and excited state can be represented as $\hat V = i\mathcal Ed_0F(t)\hat D$ and $\hat V = i\mathcal Ed_0^*F(t)\hat D^T$, correspondingly, where
\begin{equation}
\hat D = \begin{pmatrix}
 -\sqrt{\frac23}& 0 & 0& 0 \\
 0&  -\sqrt{\frac15}& 0& 0  \\
 0 & 0& -\sqrt{\frac1{10}}& 0 \\
 0 & 0& 0& -\sqrt{\frac1{30}}\\
 0 & 0 & 0 & 0 \\
  0 & 0 & 0 & 0 
\end{pmatrix},\label{DopApp}
\end{equation} 
$F(t) = p(t/T)\cos(\omega_0t)$, and $d_0$ is a reduced dipole matrix element: $d_0=\l\text{ex},J=5/2| r|\text{g},J=3/2\r$. $d_0=1$~a.~u.~is taken for simplicity. Therefore, the vectors $\mathcal C^{(1,2)}$ can be written as
\begin{align}
&\mathcal C^{(1,2)}=\mathcal E^2|d_0|^2\Biggl[\int_{-\infty}^tdt'F(t')\,\lf[\U^{(1,2)}\rt]^{-1}(t')\times\label{C12App}\\
&\quad\hat D^T\hat U_{\text{e}}(t')\int_{-\infty}^{t'}dt'F(t'')'\hat U_{\text{e}}(t'')\hat D\,\U^{(1,2)}(t'')\Biggr]
\begin{pmatrix}c\\\pm d\\d\\\pm c\end{pmatrix}\nonumber,
\end{align}

As discussed in Section \ref{SubsecEqMotionAntiferro}, $J_{x1,y1}=-J_{x2,y2}$ and $J_{z1} = J_{z2}$ during and after the action of light. Applying these relations and investigating the relations of $\nu^{(1,2)}_a$ and $\gamma^{(1,2)}_a$ to the spinors $\mathcal C^{(1,2)}$, we obtain that $\nu^{(1)}_a=\nu_a^{(2)}$ and $\gamma^{(1)}_a=\gamma_a^{(2)}$.



 \subsection{Equations of motion}

\begin{table*}[t]
\begin{equation}
\begin{array}{|c||c|c|c|c|c|}
&\hat L_x&\hat L_y&\hat M_z&\lf(\hat M_z^2+\hat L_z^2\rt)/2&\lf(\hat M_x^2+\hat L_x^2\rt)/2\\
\hline
l_{12+}'& m_{13-}&\frac32m_1-\frac32m_2-m_{13+}&-l_{12-} & -2l_{12-}& l_{12-}+\frac34 l_{23-}+l_{14-}\\
l_{12-}'&-\frac32m_1+\frac32m_2-m_{13+}&-m_{13-}&l_{12+}& 2l_{12+}& l_{12+}\frac34 l_{23+}-l_{14+}\\
l_{23+}'& -m_{13-}+m_{24-}&2m_2-2m_3  +m_{13+}-m_{24+}& -l_{23-}& 0&-l_{12-}+l_{34-}  \\
l_{23-}' & -2m_2+2m_3 +m_{13+}-m_{24+}&m_{13-}-m_{24-}&l_{23+} & 0&   -l_{12+}+l_{34+}\\
l_{34+}' & -m_{24-}&\frac32m_3-\frac32m_4+m_{24+}&-l_{34-} & 2l_{34-} & -\frac 34 l_{23-}-l_{34-}-l_{14-}\\
l_{34-}' & -\frac32m_3+\frac32m_4+m_{24+}&m_{24-}&l_{34+} & -2l_{34+}& -\frac34 l_{23+}+l_{34+}+l_{14+}\\
l_{14+}'& \frac34(m_{13-}-m_{24-})&\frac34(m_{13+}-m_{24+})&-3l_{14-}&0&\frac34 l_{12-}-\frac34 l_{34-}\\
l_{14-}'& \frac34(-m_{13+}+m_{24+})&\frac34(m_{13-}-m_{24-})&3l_{14+}&0&-\frac 34 l_{12+}+\frac34 l_{34+}\\
m_{13+}' & l_{12-}+l_{14-}-\frac34l_{23-}&l_{12+}-l_{14+}-\frac34l_{23+}&-2m_{13-}&-2m_{13-}&m_{13-}\\
m_{13-}' & -l_{12+}-l_{14+}+\frac34l_{23+}&l_{12-}-l_{14-}-\frac34l_{23-}&2m_{13+}&2m_{13+}&-\frac32 m_1-m_{13+}+\frac32m_3\\
m_{24+}'& \frac34l_{23-}-l_{14-}-l_{34-}&l_{14+}+\frac34l_{23+}-l_{34+}&-2m_{24-}&2m_{24-}&-m_{24-}\\
m_{24-}'& -\frac34l_{23+}+l_{14+}+l_{34+}&l_{14-}+\frac34l_{23-}-l_{34-}&2m_{24+}&-2m_{24+}&-\frac32 m_2+\frac32m_4+m_{24+}\\
m_1'& l_{12-}&-l_{12+}&0&0&m_{13-}\\
m_2'& l_{23-}-l_{12-}&-l_{23+}+l_{12+}&0&0&m_{24-}\\
m_3'& -l_{23-}+l_{34-}&l_{23+}-l_{34+}&0&0&-m_{13-}\\
m_4'&-l_{34-}&l_{34+}&0&0&-m_{24-}\\
\end{array}\nonumber
\end{equation}
\caption{First column: $k'$, which is equal to $-i\lf\l\lf[\hat k,\H+\Hs\rt]\rt\r$, where $\hat k$ is $\hat l_{ab\pm}$ or $\hat m_{ab\pm}$. Four left columns: $-i\lf\l\lf[\hat k,\hat O\rt]\rt\r$, where $\hat O$ denotes the operators entering $\H$: $\hat L_x$, $\hat L_y$, $\hat M_z$, $\lf(\hat M_{z}^2+\hat L_{z}^2\rt)/2$ and $\lf(\hat M_{x}^2+\hat L_{x}^2\rt)/2$.}
\label{CommutatorsApp}
\end{table*}

We solve differential equations for the expectation values
\begin{align}
&m_{12\pm}= \frac{\sqrt3}2\Bigl[\lf\l\Psi^{(1)}_{\text{g}}\lf|\hat N^{(1)}_{12\pm}\rt|\Psi^{(1)}_{\text{g}}\rt\r+\lf\l\Psi^{(2)}_{\text{g}}\lf|\hat N^{(2)}_{12\pm}\rt|\Psi^{(2)}_{\text{g}}\rt\r\Bigr],\label{ml_expval}\\
&l_{12\pm}=\frac{\sqrt3}2\Bigl[\lf\l\Psi^{(1)}_{\text{g}}\lf|\hat N^{(1)}_{12\pm}\rt|\Psi^{(1)}_{\text{g}}\rt\r-\lf\l\Psi^{(2)}_{\text{g}}\lf|\hat N^{(2)}_{12\pm}\rt|\Psi^{(2)}_{\text{g}}\rt\r\Bigr],\nonumber\\
&m_{13\pm}= \frac{\sqrt3}2\Bigl[\lf\l\Psi^{(1)}_{\text{g}}\lf|\hat N^{(1)}_{13\pm}\rt|\Psi^{(1)}_{\text{g}}\rt\r+\lf\l\Psi^{(2)}_{\text{g}}\lf|\hat N^{(2)}_{13\pm}\rt|\Psi^{(2)}_{\text{g}}\rt\r\Bigr],\nonumber\\
&l_{13\pm}=\frac{\sqrt3}2\Bigl[\lf\l\Psi^{(1)}_{\text{g}}\lf|\hat N^{(1)}_{13\pm}\rt|\Psi^{(1)}_{\text{g}}\rt\r-\lf\l\Psi^{(2)}_{\text{g}}\lf|\hat N^{(2)}_{13\pm}\rt|\Psi^{(2)}_{\text{g}}\rt\r\Bigr],\nonumber\\
&m_{34\pm}= \frac{\sqrt3}2\Bigl[\lf\l\Psi^{(1)}_{\text{g}}\lf|\hat N^{(1)}_{34\pm}\rt|\Psi^{(1)}_{\text{g}}\rt\r+\lf\l\Psi^{(2)}_{\text{g}}\lf|\hat N^{(2)}_{34\pm}\rt|\Psi^{(2)}_{\text{g}}\rt\r\Bigr],\nonumber\\
&l_{34\pm}=\frac{\sqrt3}2\Bigl[\lf\l\Psi^{(1)}_{\text{g}}\lf|\hat N^{(1)}_{34\pm}\rt|\Psi^{(1)}_{\text{g}}\rt\r-\lf\l\Psi^{(2)}_{\text{g}}\lf|\hat N^{(2)}_{34\pm}\rt|\Psi^{(2)}_{\text{g}}\rt\r\Bigr],\nonumber\\
&m_{24\pm}= \frac{\sqrt3}2\Bigl[\lf\l\Psi^{(1)}_{\text{g}}\lf|\hat N^{(1)}_{24\pm}\rt|\Psi^{(1)}_{\text{g}}\rt\r+\lf\l\Psi^{(2)}_{\text{g}}\lf|\hat N^{(2)}_{24\pm}\rt|\Psi^{(2)}_{\text{g}}\rt\r\Bigr],\nonumber\\
&l_{24\pm}=\frac{\sqrt3}2\Bigl[\lf\l\Psi^{(1)}_{\text{g}}\lf|\hat N^{(1)}_{24\pm}\rt|\Psi^{(1)}_{\text{g}}\rt\r-\lf\l\Psi^{(2)}_{\text{g}}\lf|\hat N^{(2)}_{24\pm}\rt|\Psi^{(2)}_{\text{g}}\rt\r\Bigr],\nonumber\\
&m_{14\pm}= \frac{3}4\Bigl[\lf\l\Psi^{(1)}_{\text{g}}\lf|\hat N^{(1)}_{14\pm}\rt|\Psi^{(1)}_{\text{g}}\rt\r+\lf\l\Psi^{(2)}_{\text{g}}\lf|\hat N^{(2)}_{14\pm}\rt|\Psi^{(2)}_{\text{g}}\rt\r\Bigr],\nonumber\\
&l_{14\pm}=\frac{3}4\Bigl[\lf\l\Psi^{(1)}_{\text{g}}\lf|\hat N^{(1)}_{14\pm}\rt|\Psi^{(1)}_{\text{g}}\rt\r-\lf\l\Psi^{(2)}_{\text{g}}\lf|\hat N^{(2)}_{14\pm}\rt|\Psi^{(2)}_{\text{g}}\rt\r\Bigr],\nonumber\\
&m_{23\pm}= \Bigl[\lf\l\Psi^{(1)}_{\text{g}}\lf|\hat N^{(1)}_{23\pm}\rt|\Psi^{(1)}_{\text{g}}\rt\r+\lf\l\Psi^{(2)}_{\text{g}}\lf|\hat N^{(2)}_{23\pm}\rt|\Psi^{(2)}_{\text{g}}\rt\r\Bigr],\nonumber\\
&l_{23\pm}=\Bigl[\lf\l\Psi^{(1)}_{\text{g}}\lf|\hat N^{(1)}_{23\pm}\rt|\Psi^{(1)}_{\text{g}}\rt\r-\lf\l\Psi^{(2)}_{\text{g}}\lf|\hat N^{(2)}_{23\pm}\rt|\Psi^{(2)}_{\text{g}}\rt\r\Bigr],\nonumber\\
&m_{a}= \Bigl[\lf\l\Psi^{(1)}_{\text{g}}\lf|\hat N^{(1)}_{a\pm}\rt|\Psi^{(1)}_{\text{g}}\rt\r+\lf\l\Psi^{(2)}_{\text{g}}\lf|\hat N^{(2)}_{a\pm}\rt|\Psi^{(2)}_{\text{g}}\rt\r\Bigr],\nonumber\\
&l_{a\pm}=\Bigl[\lf\l\Psi^{(1)}_{\text{g}}\lf|\hat N^{(1)}_{a\pm}\rt|\Psi^{(1)}_{\text{g}}\rt\r-\lf\l\Psi^{(2)}_{\text{g}}\lf|\hat N^{(2)}_{a\pm}\rt|\Psi^{(2)}_{\text{g}}\rt\r\Bigr],\nonumber
\end{align}
where $a$ is 1,2,3 or 4. The initial values of $m_{ab\pm}(0)$ and $l_{ab\pm}(0)$ are given by substituting $\Psi^{(1,2)}_{\text{g}}(0)=\Psi^{(1)}_{0}$. We obtain that $m_{12\pm}(0)=0$, $m_{23\pm}(0)=0$, $m_{34\pm}(0)=0$, $m_{14\pm}(0)=0$, $l_{13\pm}(0)=0$, $l_{24\pm}(0)=0$ and $l_a(0)=0$ for $a=1\dotsc4$ and these variables remain zero at any time. Applying that $l_a(0)=0$, the equations of motion can be written as
\begin{align}
m_{ab\pm}'  = &\left(-\sum_k\nu_km_k+\nu_a+\nu_b\right)m_{ab\pm}\label{EqMot_for_mApp} \\
&\pm(\gamma_a-\gamma_b) m_{ab\mp}-i\lf\l\lf[\hat m_{ab\pm}, \H\rt]\rt\r\nonumber\\
l_{ab\pm}'  = &\left(-\sum_k\nu_km_k+\nu_a+\nu_b\right)l_{ab\pm}\label{EqMot_for_lApp}\\
&\pm(\gamma_a-\gamma_b) l_{ab\mp}-i\lf\l\lf[\hat l_{ab\pm}, \H\rt]\rt\r.\nonumber
\end{align}
The relations $m_{12\pm}(t)=0$, $m_{23\pm}(t)=0$ and $m_{34\pm}(t)=0$ lead to $M_x=M_y=0$, and $l_a(0)=0$ for $a=1\dotsc4$ leads to $L_z=0$. Thus, $\H$ can be represented as
\begin{align}
&\H = \frac{\J}{2}\left(-L_x\hat L_x-L_y\hat L_y+M_z\hat M_z\right)+\hat{\mathcal H}_{\text{cr}z(\text{cr}x)}\label{H0AntiferroML},\\
&\hat{\mathcal H}_{\text{cr}z(\text{cr}x)}=3\Delta\left(\frac{\hat M_{z(x)}^2+\hat L_{z(x)}^2}{2}+\frac{\hat J_1^2+\hat J_2^2}3\right).\nonumber
\end{align}

Table \ref{CommutatorsApp} shows the time derivatives of all involved variables $m_{ab\pm}$ and $l_{ab\pm}$ and corresponding expressions for $-i\l[\hat m_{ab\pm}, \hat O]\r$ and $-i\l[\hat l_{ab\pm}, \hat O]\r$, where $\hat O$ denotes the operators entering $\H$: $\hat L_x$, $\hat L_y$, $\hat M_z$, $(\hat M_z^2+\hat L_z^2)/2$ and $-(\hat M_x^2+\hat L_x^2)/2$. For instance, it follows from Eqs.~(\ref{EqMot_for_lApp}) and Table \ref{CommutatorsApp} that, in the case of $\Hcr=\Hcrz$,
\begin{align}
&l_{12+}'=\left(-\sum_{a=1}^4\nu_am_a+\nu_1+\nu_2\right)l_{12+}+(\gamma_1-\gamma_2) l_{12-}\\
&\quad-\frac{\J}{2} \Bigl[L_xm_{13-}+L_y\left(\frac32m_1-\frac32m_2-m_{13+}\right)\nonumber\\
&\qquad\qquad+M_zl_{12-}\Bigr]-\DD l_{12-}\nonumber.
\end{align}

\end{appendix}


%

\end{thebibliography}

\end{document}